\documentclass[aps,amssymb,prl,twocolumn]{revtex4}
\usepackage{graphicx} \usepackage{epsfig} \usepackage{latexsym}
\usepackage{color}

\newcommand{\cut}{c_{i-1}}

\begin{document}
\bibliographystyle{prsty}
\title{Highly Optimized Tolerance and Power Laws in Dense and
Sparse
Resource Regimes}

\author{M. Manning$^{1}$, J. M. Carlson$^{1}$, J. Doyle$^{2}$}
\affiliation{
$^{(1)}$ Department of Physics, University of California, Santa
Barbara, California 93106, U.S.A. \\
$^{(2)}$ Department of Control and Dynamical Systems and
Electrical Engineering, California Institute of Technology,
Pasadena, California 91125, U.S.A.}

\date{\today}

\begin{abstract}
Power law  cumulative frequency $(P)$ vs.~event size $(l)$
distributions $P(\geq l)\sim l^{-\alpha}$ are frequently cited
as evidence for complexity and serve as a starting point
for linking theoretical models and mechanisms with observed
data. Systems  exhibiting this behavior present fundamental
mathematical challenges in probability and statistics.
The broad span of length and time
scales associated with heavy tailed  processes 
often require special sensitivity to 
distinctions between discrete and continuous phenomena.
A discrete Highly Optimized Tolerance (HOT) model,
referred to as the Probability, Loss, Resource (PLR) model,
gives the exponent $\alpha=1/d$ as a function of the
dimension $d$ of the underlying substrate in the sparse
resource regime. This agrees well with data for wildfires,
web file sizes, and electric power outages. However,
another HOT model, based on a continuous (dense)
distribution of resources, predicts $\alpha= 1+ 1/d $. In
this paper we describe and analyze a third model, the cuts
model, which exhibits both behaviors but in different
regimes. We use the cuts model to show all three models
agree in the dense resource  limit. In the sparse resource
regime, the continuum model breaks down, but in
this case, the cuts and PLR models are described by the
same exponent.
\end{abstract}
\maketitle



\section{I. Introduction}

In this paper we analyze a family of abstract, mathematical
models which have been used to illustrate {\em Highly
Optimized Tolerance} (HOT)~\cite{PRE1, PRL1,PLR,PNAS,DDOF,COLD}, a
mechanism for complexity based on robustness tradeoffs in
systems subject to uncertain environments. HOT systems
abound in nature and modern technology, and are complex and
highly structured. They arrive at \lq\lq optimized" or
\lq\lq organized" states through deliberate design or
biological evolution, and exhibit robust, yet fragile (RYF)
characteristics, the essence of HOT. That is, they are
robust to normal or common perturbations,  yet may be
extremely fragile to rare perturbations or design flaws,
even if the perturbations are small and seemingly
innocuous.

Recently, HOT has been investigated in the context of a
variety of specific applications, including the
Internet~\cite{INFOCOM,Heuristic}, the Electric Power
Grid~\cite{CORNELL}, Wildfires~\cite{fire}, and Biological
Networks~\cite{bionetworks1, bionetworks2, heatshock, bionetworks3, bionetworks4, epidemic, dynamics}. Typically,
these studies involve a combination of simple abstract,
analytically tractable representations, which focus on
fundamental tradeoffs and derivations of the power laws,
with detailed, high-resolution simulation models, aimed at
pinpointing specific system and  model fragilities. Here we
focus specifically on the abstract models which have been
used to describe HOT. We compare discrete and continuum
models in a common framework, and clarify the
approximations that are made and the ranges of 
applicability of the models. This forces us to address
certain fundamental issues in probability and statistics, including 
distinctions between discrete and continuous distributions, and 
properties associated with mixtures of distributions.

One key success of HOT is to offer an alternative
perspective on the origins and ubiquity of complexity, and
particularly power laws.  Mathematically, heavy tailed
distributions  (e.g.~power laws) often require special care 
because of the broad range of spatial and temporal scales over 
which data is sampled \cite{fellerII,johnson,StableLaws,
Mandelbrot97,Mitzenmacher03,wsc04}. In many cases,  conventional
assumptions and methodologies associated with modeling and 
data analysis are misleading and/or break down. One of the
goals of this paper is to illustrate how such problems can
arise, and to approach them in a manner which is mathematically
rigorous.

HOT has been compared to earlier
work emphasizing emergent complexity, where power laws
arise from minimal tuning, on an otherwise random
substrate.  In emergent complexity power laws are
associated with fractals and self-similarity \cite{SPrefs,SOC_first}.
In many studies, HOT illustrates the differences between
organized and emergent complexity by using percolation
forest fire models from physics \cite{Malamud,FFrefs,FFrefs2}, but
including a minimal form of optimization (intended to
capture design or evolution) and robustness tradeoffs
\cite{PRE1,PRL1,DDOF,COLD,Z_PNAS}.  This produces power
laws (in better agreement with data) that arise from highly
organized and self-dissimilar structures, the opposite of
self-similarity.

All of the abstract HOT models follow the same basic
mechanistic description involving optimization of tradeoffs
in an uncertain environment. Each begins with a
$d$-dimensional substrate representing the system. Each
event (e.g., a power outage or fire) is triggered by  some
small perturbation or spark (typically chosen from a
nonuniform distribution) which initiates a cascading
failure, resulting in loss of some portion of  the
substrate. All of the models considered here assume the
loss (or cost) associated with an event scales linearly
with the event size. Alternative cost functions give power
laws in cost, not necessarily raw event size ~\cite{PRE1}.
Thus cost functions that heavily weight large events
can lead to truncation of the power laws~\cite{COLD}. 

In HOT, resources  are allocated to create barriers limiting
propagation of the cascading failure events in a manner
which  optimizes  the cost function   (minimizing loss or
maximizing yield). There is a limited number of resources
available, and this constraint is modeled in one of two
ways. The first method places a fixed limit on the total
resources available. The second weights resource use
alongside other costs or losses, which are associated with the
events themselves, by including an explicit
resource term in the cost function.   Here the key issue is
to account explicitly for resource use.  The specific form
of the constraint does not play a significant role in
determining the size distribution.

In HOT,  optimization of the resource allocations subject
to the constraint represents design and/or evolutionary
tradeoffs in systems faced with  a spectrum of
disturbances. Because resources are constrained  and often
sparse or expensive, optimal solutions make efficient use
of the resources available, resulting in HOT states
characterized by  structured, compact, $d$-dimensional
regions surrounded by $(d\!-\!1)$-dimensional barriers. In
addition, for  a broad class of distributions of
disturbances (e.g. Gaussian, exponential, and Cauchy),
minimization of the average loss results in heavy-tailed,
power law distributions in the sizes of the events. Newman
{\it et al.}~\cite{COLD} emphasize however that the
specific exponents characterizing the decay of the power
law distribution in HOT models can be different.

In this paper we  focus on three models for HOT which are
among the simplest, and most analytically tractable
examples. Table~\ref{tbl:compare} summarizes their basic
properties, which will be described in detail in the
following sections. In each case,
\begin{itemize}
{ \item Probability $p$: represents uncertainty in the
environment.
\item Loss $l$: represents the volume or size associated with an
individual event, which is directly proportional to the cost of
that
event.
\item Resources $r$: provide mechanisms to limit losses.
\item Constraints: are imposed on the resources.
\item Optimization: of the resource assignments subject to
constraints leads to the HOT state.
\item Power Laws: in the cumulative event distributions,
$P(\geq l)$ vs. $l$, are characteristic
of these optimal solutions.}
\end{itemize}
All of these models are motivated by studies of the HOT
version of the percolation forest fire model
\cite{PRE1,PRL1}. The most well studied of these are the
continuum model~\cite{PRE1}, generalized by Newman et.
al.~\cite{COLD}, and the Probability  Loss Resource
(PLR)~\cite{PLR} model. Their abstractions differ in
subtle, yet important ways, leading to differences in the
predictions. The
continuum model aims to describe the continuum limit of the
HOT percolation forest fire model~\cite{PRE1}, building on
lattice models from statistical
physics~\cite{Percolation,FFrefs}, and introducing a
mean-field-like analysis of the continuum limit. In the
continuum model all aspects of the system are described as
smoothly varying functions on the substrate. The PLR model
is a generalization of Shannon Source Coding
Theory~\cite{Shannon} from Information
Theory~\cite{InfoTheory}, perhaps the simplest design model
in engineering. The PLR model begins with discrete event
categories $i$, each of which has a characteristic
probability, resource allocation, and resulting loss. Like
the continuum model, the cuts model~\cite{PRE1,INFOCOM} can
be thought of as the limiting description of a lattice
model as the lattice size becomes infinite. The cuts model
represents space continuously (like the continuum model)
but divides it into discrete regions (like the PLR model)
using sharp barriers, i.e.~cuts.

\begin{table}[h]
\begin{center}

\begin{tabular}{||p{2cm}|p{2cm}|p{2.2cm}|p{2cm}||} \hline
            & {\bf Continuum}         & {\bf PLR}   & {\bf Cuts}
\\ \hline \hline

{\small{\bf Probability} }& Continuous $p({\bf x})$ & Discrete
$p_i$ & Continuous $p(x)$ cut into $p_i$
\\ \hline

{\small {\bf Resources}} & Continuous $r(x)$ & Discrete $r_i$
& Discrete cuts \\ \hline

{\small {\bf Constraint}} & Resource cost $R=\int r(x)dx$ &
Resource
limit $\sum r_i\le R$
& $N$ cuts  \\ \hline

{\small {\bf Losses}} & Continuous $l(x)$ & Discrete $l_i$
& {\bf Discrete} $l_i$ \\ \hline

{\small {\bf Optimize} }  &  $Y \!=\! 1\! - \!\int p l\! -
\!R$ &
 $Y\! =\! 1
\!-\!\sum p_il_i$ & $Y\!=\! 1\! -\! \sum p
l$  \\ \hline

{\small {\bf Power law}} $P(\geq\!l)$ vs. $l$ &
\hspace{.5cm}$l^{-(1 +
1/d)}$ &\hspace{.5cm}
$l^{-1/d}$  &  $l^{-2}$ as $l \rightarrow 0$, $l^{-1}$ as $l
\rightarrow \infty$ ($d=1$) \\ \hline

\end{tabular}
\caption{The HOT continuum, PLR, and cuts models predict
power laws based  on optimal allocation of limited
resources to minimize loss in an uncertain environment.
Different assumptions in the continuum and PLR models lead
to different exponents in the dense and sparse resource
regimes, both of which can arise as (opposite) limits of
the cuts model. The PLR model can be extended to the dense
resource limit (Section V), where it agrees with the
continuum and cuts model. The PLR cumulative probability
$P(\geq l)$  assumes densely sampled data (Section III).
To increase readability, constant
factors are set to unity in the equations for Yield $Y$
which is optimized.} \label{tbl:compare}
\end{center}
\end{table}

A key distinction between the models is their predictions
for power law exponents. The continuum model predicts a
power law in $P(\geq l)$ with exponent $\alpha=1/d+1$, while
PLR predicts an exponent of $\alpha=1/d$ for the same
distribution of sparks (assuming densely sampled data).
We show the first two models match
solutions in different limits of the cuts model with an
exponential distribution of sparks.
In $d=1$
the cuts model predicts an exponent of $\alpha=2$ in the
limit of small events and $\alpha=1$ in the large-event
limit. Thus the cuts model captures the power laws
predicted by the other two models in the limit of small
(continuum) and large (PLR) events.

Analysis of the cuts model provides a unifying picture for 
all the models, and a concrete illustration of how certain key 
approximations made in the first two models can break down.  
We show that when the PLR and cuts models have sufficiently similar 
assumptions, their results agree as expected.  In the dense 
resource regime limit (described by the continuum model), all 
three models agree.  The cuts model also illustrates how the exponent 
describing small events departs from this dense resource limit as the 
density of resources and barriers becomes lower.

In the remaining sections of this paper, we first summarize
results for the continuum (Section II) and PLR (Section
III) models, with special attention to derivation of the
power laws, and specific features which will be useful for
comparing  models. We also discuss mathematical subtleties
which can arise in taking continuum limits in systems
with sharp barriers, as well as mathematical issues 
which can arise in comparing continuum vs.~discrete models
and distributions,
and distributions composed of  finite mixtures of
probability distributions.  The next three
sections comprise the bulk of the new analytical results in
this paper. In Section IV we review and extend the cuts
model and in Section V we compare it to the other models.
We show that the event size distribution for the PLR and cuts 
model agrees when their assumptions are forced to be similar.  
They also both agree with the continuum model in the limit of 
dense resources and small event sizes. 
For the cuts model with 
a power law distribution of sparks, the small event limit is 
described by a power law in which the exponent depends on 
the distribution of sparks, ranging from the limiting value 
of $\alpha=2$ (which we obtain for an exponential spark 
density) for an infinitely steep power law, to $\alpha=1$ 
(sparse resource limit, and in agreement with PLR) 
when $p(x)\sim 1/x$. Furthermore, for both exponential and 
power law distributions of sparks, we find that the event 
size distribution for the cuts model agrees with the PLR model 
in the limit of large event sizes, where the distribution is 
clearly discrete. In this case the agreement between models
depends on the assumption of a 
sufficiently well sampled data set, which would only arise 
in the cuts and PLR models due to mixtures.  
In Section VI we return to the
original HOT lattice model, and illustrate a  subtle
pathology  which arises in the continuum limit of the
lattice model in the absence of an explicit resource cost
or constraint. In Section VII we conclude with a discussion
of our results, and the relevance of the different resource
regimes in the context of  observed data.

\section{II. Continuum model}

The continuum model was first suggested as  an approximate
limiting description of  the HOT forest fire percolation
lattice model by Carlson and Doyle~\cite{PRE1}. It was
later studied alongside large lattice model simulations by
Newman et al.~\cite{COLD}. The definition of the model most
conveniently begins with the lattice model, which we will
return to in Section VI. Strictly speaking, the continuum
model is an  approximation to the lattice model based on
scaling arguments. It captures the power laws observed in
the HOT lattice model in the limit of large, finite lattice
sizes, and allows the size distribution to be calculated
analytically.

\begin{figure}
\resizebox{!}{.18\textwidth}{{\includegraphics{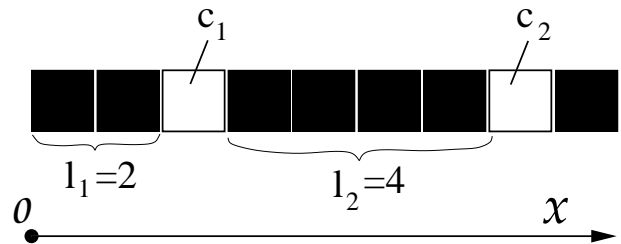}}}
\caption{Sample configuration of the percolation forest fire
model in
$d=1$. Occupied sites (black) correspond to
trees, and vacant sites (white) correspond to firebreaks. When a
spark hits an occupied site it  burns all trees
in the connected cluster (labeled $l_i$) of occupied sites
containing
the initiating site.
Fire terminates in each direction upon encountering a firebreak,
or
{\em cut},  labeled $c_i$.}
\label{1d_lattice}
\end{figure}

Consider a $d-$dimensional space, with positions labeled by
the $d-$dimensional vector ${\bf x}$ (these are discrete
sites on a hypercubic lattice, each labeled by $d$ integer
indices $i,j,k...$, with ${\bf x}=(i/N, j/N, k/N....)$,
where $N$ is the number of sites along each axis of the
lattice). In the percolation lattice model, each position
(site) is either occupied by a tree, or vacant (firebreak).
Environmental uncertainty is represented by the probability
$p({\bf x})$ that a spark lands at site ${\bf x}$. A spark
ignites a fire that spreads throughout the nearest neighbor
connected cluster of trees in all  $d$ directions, but
terminates at firebreaks. The resulting fire size is the
total  number of sites in the burned patch, $l({\bf x})$,
and the value of $l({\bf x})$ is clearly constant within
each  contiguous patch. A sample  lattice  configuration in
the special case $d=1$ is illustrated in
Figure~\ref{1d_lattice}. Occupied sites (black) are trees
and unoccupied sites (white) are firebreaks. Event sizes
$l_i$ correspond to the number of occupied spaces between
firebreaks, or {\em cuts}, labeled $c_i$.

HOT configurations optimize the layout of vacant and
occupied sites  to maximize yield $Y$, defined to be the
average number of occupied sites which remain after a
single spark lands on the lattice (averaging over the spark
distribution $p({\bf x})$). For small lattices, it is
possible to compute the globally optimal
solution~\cite{DDOF}.  However, for large lattices the
solution becomes computationally intractable (and not
especially informative). Instead, a wide variety of
constrained optimization schemes have been
investigated~\cite{DDOF, PRE1, PRL1, epidemic, dynamics},
all leading to similar results. Firebreaks are concentrated
in regions of high spark probability, so that only small
fires occur in regions of the lattice where sparks are
common, while large fires occur in regions where sparks are
rare.

The specialized, patterned HOT configurations reflect
patterns in the perturbing environment. This is in sharp
contrast to the traditional forest fire percolation model
studied in statistical physics~\cite{Percolation}, where
configurations are essentially random, aside from a tuned,
or \lq\lq self-organized" average critical density
\cite{SPrefs,SOC_first,Malamud,FFrefs,FFrefs2}. The contrasts between
the HOT and self-organized critical lattice models are
discussed in detail in \cite{PRE1, PRL1,PLR,PNAS,DDOF}, and will
not be our emphasis here. 

The HOT lattice model was the
first model introduced to illustrate the HOT mechanism, and
is pedagogically useful in illustrating the emergence of
$(d-1)$-barriers on the $d-$dimensional substrate, as well
as the high concentrations of barriers in regions where
perturbations are common. All of the other models
considered here retain these key features, but each
explicitly accounts for the cost
of resources in a different way. More importantly, each makes different
approximations in representing continuum versus discrete
spatial features of the lattice model, which lead to the
different predictions for the event size distribution.

In the continuum model  the integer $i/N$ components of the
$d-$dimensional vector positions ${\bf x}$ are replaced in
the limit $N\to\infty$ by real  valued components. The
occupied (tree) and vacant (firebreak) lattice sites are
replaced by a resource density $r({\bf x})$, representing
the local density of firebreaks. A function $l({\bf x})$
represents the size of the loss which occurs when a spark
lands at position ${\bf x}$. A key approximation relative
to the original lattice model is clearly made in the
continuum model, which represents $r({\bf x})$ and $l({\bf
x})$ as  continuous functions. The idea is to use a scaling
relation, motivated by the lattice model, to mimic the
manner in which higher resource densities lead to smaller
fires in a given region, without accounting in detail for
the specific configuration.

To derive the distribution of fire sizes for the continuum
model, we follow the elegant derivation of Newman et
al.~\cite{COLD}. The size of a firebreak surrounding a
given patch $l({\bf x})$ is:
\begin{equation}
r({\bf x}) = gd l({\bf x})^{(d-1)/d},
\label{r_of_x}
\end{equation}
where $g$ is a geometric factor of order unity that depends
on the shape of the patch. It is in Eq.~(\ref{r_of_x}) that
the dimensional relationship between resource and loss is
captured. In the  continuum model  the total resource use
is given by
\begin{equation}
R=\int r({\bf x}) d{\bf x},
\label{R_continuum}
\end{equation}
where the integral is over the $d-$dimensional substrate.
In the continuum model, this cost enters explicitly into
the yield function. Normalizing $Y$ by the total volume of
the substrate (i.e. $Y=1$ corresponds to a fully occupied
forest, with no fires or firebreaks), and averaging over
the distribution of sparks $p({\bf x})$, we write the
expected yield as
\begin{equation}
\label{cont_yield}
  Y = 1 - c\int p({\bf x}) l({\bf x})d{\bf x} - a R
\end{equation}
where $c$ is the cost per unit area (or generally,
$d$-dimensional volume) of forest, and $a$ is the cost per
unit length (or $(d\!-\!1)$-dimensional volume) of
firebreaks. This yield function is motivated by tradeoffs
inherent in the original lattice model, where the
resources are empty sites, and the cost of firebreaks is
the yield penalty in  initial density associated with
creation of vacancies. However, unlike the lattice model,
it includes a nonvanishing resource term explicitly in the
yield function, and allows the constants $a$ and $c$ to
scale differently with dimension $d$. This fortuitously
omits a pathology which results from the difference in
scaling between the compact, $d-$dimensional clusters of
trees, and the $(d-1)$-dimensional firebreaks which arises
in the lattice model as $N\to\infty$. We discuss this in
more detail in Section VI.

The optimal allocation of resources $r({\bf x})$ maximizes
the expected yield. Optimizing resources is equivalent to
optimizing over event sizes because they are explicitly related
via Eq.~(\ref{r_of_x}).  To obtain the solution, we assume
that $l({\bf x})$ is a continuous function of the ignition
site ${\bf x}$, and  set the functional derivative $\delta
Y/ \delta l({\bf x})$ equal to zero. This  leads to
\begin{equation}
\label{cont_l}
  l({\bf x}) = C p({\bf x})^{-d/(d+1)}
\end{equation}
where C is a constant that depends on $a$,$c$, and $g$.

\begin{figure}
\resizebox{!}{.22\textwidth}{{\includegraphics{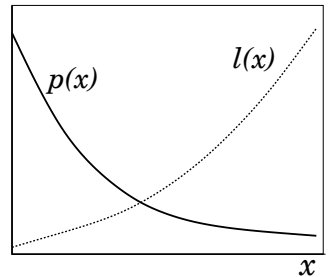}}
}
\caption{Schematic solution of the continuum model in $d=1$.
Small
event sizes
$l({\bf x})$ are associated with positions ${\bf x}$ of high
spark probability $p({\bf x})$. Eliminating ${\bf x}$ and
integrating
leads to the event size distribution
$P(\geq l)$ as described in the text. A key distinguishing
feature of the continuum model is that
the event size function $l({\bf x})$ is {\it a priori}
a continuous function of ${\bf x}$.}
\label{l_of_x_continuum}
\end{figure}

A schematic solution in $d=1$ is illustrated in Figure
\ref{l_of_x_continuum}. It is important to note that the
continuum model  departs from the original lattice model in
representing $l({\bf x})$ as a continuous function of ${\bf
x}$. For a given configuration in the lattice model,
$l({\bf x})$  assumes a constant, finite value for each
contiguous cluster of occupied sites. Therefore $l({\bf
x})$  in that case, is piecewise constant. The continuum
model represents $ l({\bf x})$ as continuous over the
entire space, leading to Eq.~(\ref{cont_l}). It is the only
one of the models we consider which builds in this assumption.

It is also possible to calculate the probability density
$\rho(l)$ of fire sizes for the continuum model. Again,
assuming $l({\bf x})$ is continuous we
obtain~\cite{PRE1,COLD}:
\begin{eqnarray}
\label{cont_ps}
 \rho(l) &= p({\bf x}) \frac{d^{d}x}{dl} = p({\bf
x})\frac{d^{d}x}{dp}\frac{dp}{dl} \nonumber \\
&= \left[ p({\bf
x})\frac{d^{d}x}{dp} \right] C^{\prime} l^{-(2+1/d)}
\end{eqnarray}
 where $C^{\prime}$ is a constant that depends on $d$,$c$,$a$,
and
$g$. Newman {\it et al.} thoroughly investigated the behavior of
$\rho(l)$ and found that the scaling  behavior is  dominated by
the factor of $l^{-(2+1/d)}$, while the factor $p({\bf
x})\frac{d^{d}x}{dp}$ generates at most logarithmic corrections
for a broad class of probability distributions $p({\bf
x})$~\cite{COLD}.

Since the probability density  $\rho(l)$ is continuous,
the cumulative distribution of events of size greater than or
equal
to $l$,
$P~(~\geq~l~)$, is
proportional to $l^{-(1+1/d)}$. Therefore, for a one dimensional
substrate, the continuum model predicts a slope of
$\alpha=2$ for the cumulative distribution of events.
Table~\ref{tbl:compare} summarizes the properties of this model.

\section{III. Probability Loss Resource Model (PLR)}

The PLR (Probability Loss Resource) HOT model is a
generalization of Shannon Source Coding Theory for data
compression \cite{Shannon}, the simplest, most elegant
design theory in engineering. It is the simplest model
illustrating HOT \cite{PLR}, and is based on optimal
allocation of limited resources, with an explicit, fixed
cap on the total resources available. It retains a
dimension-dependent relationship between  resources
($(d-1)$-dimensions) and  loss ($d$-dimensions), but
otherwise replaces the explicit spatial variable ${\bf x}$
with a more abstract notion of event categories $i$. The
idea is to group similar conditions, from the common to the
rare, into a category, represented by  the relative
probabilities $p_i$.

The PLR objective is to allocate resources in a manner
which maximizes yield $Y$ averaged over a spectrum of
possible events:
\begin{equation}
Y  =  1-c \sum p_i l_i
| \ l_i=f(r_i), \ \sum r_i \le R .
\label{Jdef}
\end{equation}
Here $c$ is a constant, and $i$, $1\le i\le N$, indexes the
finite and discrete set of probabilities $p_i$, assumed to
be in descending order, with corresponding loss $l_i$.
Normalized, the cumulative $P(\geq l_i)=\sum_{j\geq i} p_j$ is the rank
order divided by the total number of events in a data set,
from which corresponding values of $p_i$ may be deduced. We
will interpret the $p_i$ as probabilities, so $Y$ is
average yield, but in general the $p_i$ could be any
weights assigned to create a cost function.

The probability $p_i$ of each category is fixed, and a
total resource  allocation $r_i$ is made to the event
category $i$, resulting in events of size $l_i$ for the
category (i.e. $r_i$ is the total resource allocation to
all the events in event category $i$). The $r_i$ are chosen
to minimize the average event size, averaged over the
spectrum of possible conditions $\{i\}$. The only
interaction between events is that the sum of all resources
is limited by $\sum r_i \le R$. This means that any
reasonable design will devote more resources to the
categories of common events so that they yield small
losses, leaving relatively few resources for rare events.

Unlike explicitly spatial lattice models, the PLR model
presumes a mean-field-like independence of events.
However, a lattice abstraction (which should not be
interpreted as a literal gridding of the forest) can be
used to derive the relationship $l_i=f(r_i)$ between
resource allocation and loss for the event categories
$\{i\}$. Imagine a large, finite $d-$dimensional lattice
which is an abstraction of a space representing a single
condition category $i$. The lattice is of length $L$ on
each side, and the total volume $L^d$ serves as the large
scale cutoff, i.e. the size of the largest possible event.
The value of $p_i$ is the total probability of hitting any
part of the lattice for category $i$, and the probability
of hitting any one of the cells within category $i$ is
equal. Resources $r_i$ represent the total allocation of
vacant sites within the $i^{th}$ category.

Because the spark distribution $p_i$ is uniform within each
category, the optimal use of resources (vacant sites)
defines a collection of equally spaced $(d-1)$-dimensional
surfaces, one lattice spacing wide in the remaining
dimension, on an otherwise occupied lattice. This defines a
set of compact, contiguous cells, all of equal size $l_i$,
for category $i$. For example, in $d=1$, the barriers
correspond to a single unoccupied site between contiguous
occupied sites of equal length. This is similar to the
lattice shown in Figure~\ref{1d_lattice}, except the
occupied regions $l_1,l_2, \ldots$ would all have the same
length.

Suppose a resource allocation of size $r_i=\sum_d L\xi$
(number of vacancies) is made to category $i$,  arranged as
$\xi$  equally spaced cuts, spanning the full length of the
lattice $L$ in each dimension $d$. Then the event size
$l_i$ for category $i$ is $l_i=((L/\xi)-1)^d$. Eliminating
$\xi$ yields a relationship between event size $l_i$ and
the resource allocation $r_i$, which scales like $l_i\sim
{r_i}^{-d}$. Here $L$ is simply  the constant subregion
lattice length scale, and the key result is the dimensional
relationship between resource allocation (to the event
category as a whole) and the corresponding characteristic
loss size for that category.

\begin{figure}
\resizebox{!}{.22\textwidth}{{\includegraphics{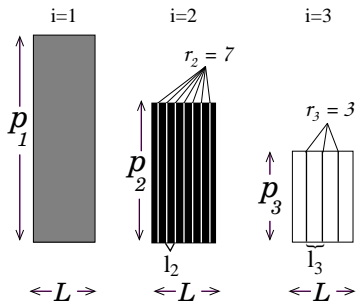}}}
\caption{\label{PLR_nc} Resources allocated to event
category $i$ in the PLR model in $d=1$ divide a region of
fixed length $L$ (horizontal axis) into events of equal
length $l_i$, characteristic of the category. In the
optimal solution, regions of high probability (vertical
axis) are allocated more resources, resulting in smaller
events.}
\end{figure}

This process is illustrated for $d=1$ in Figure
\ref{PLR_nc}. Three event categories ($i=1,2,3$) are
shown, in order  of descending probability $p_1>p_2>p_3.$
Here the constant subregion size $L$ is the identical horizontal
length of the line segment associated with each region. The
vertical height of each box reflects the probabilities
$p_i$ (and is not related to any spatial dimension or
length scale).  Resources $r_i$ are allocated to each
region, with $r_1>r_2>r_3$, and divide each region into
line segments $l_i$ equal size, with $l_1<l_2<l_3$.

Incorporating a cutoff at small event sizes, and
normalizing so that $0\le r_i\le 1$ with f(1)=0, in
$d-$dimensions we write
\begin{equation}
\label{resource}
f(r_i)={\gamma \over d}({r_i}^{-d}-1),\ \ \ \ d>0,
\end{equation}
which incorporates the scaling determined above, and
uniquely determines $f(r_i)$ up to the parameter $\gamma$.
As in the original Shannon Theory, we relax the constraint that
the $r_i$ take integer values.
This is an extremely simple and tractable model with
essentially only one parameter, the dimension of the
substrate, where events are characterized by
$d-$dimensional, compact regions, enclosed by
$(d-1)$-dimensional perimeters.

Given a fixed resource budget $R$, the goal outlined in
Eq.~(\ref{Jdef}) is to optimize the division of resources
$r_i$ to maximize yield, by minimizing the expected loss
$\sum_i{p_i l_i}$, subject to the resource vs. loss
relationship in Eq.~(\ref{resource}).
This is  accomplished using standard constrained
optimization methods (Lagrange multipliers). Setting the
gradient of $\lambda (\sum r_i-R) +\sum p_i f(r_i)$ equal
to zero yields $-p_i{f}'(r_i)=\lambda$, which equalizes the
expected marginal loss and can be solved for the $r_i$.
Then the optimal $\lambda$ saturates the resource
constraint with $\sum r_i = R$, $r_i \leq 1$, yielding
\begin{equation}
r_i = R p_i^\frac{1}{1+d}
\left( \sum_j p_j^\frac{1}{1+d} \right)^{-1},
\label{r}
\end{equation}
so that
\begin{equation}
l_i =
{{\gamma}\over{d}}
\left[
\left( R p_i^\frac{1}{1+d} \right)^{-d}
\left( \sum_j p_j^\frac{1}{1+d} \right)^{d}
-1 \right],
\label{l}
\end{equation}
and
\begin{equation}
J= {d}^{-1}\left[ R^{-d} \left(\sum_i p_i^\frac{1}{1+d}
\right)^{1+d} - \sum_i p_i \right].
\label{Jopt}
\end{equation}
Inverting (\ref{l})  yields a relationship between the
event type and corresponding probability:
\begin{equation}
   {p}_i (l_i)  =    {1\over \alpha}(C+ l_i)^{-(1+\alpha)}
\label{pdist}
\end{equation}
where $\alpha=1/d$ and $C$ is a constant (which depends on
$\gamma$, $d$, and $R$ in Eq.~(\ref{resource})) which sets
the small size scale in the resource vs. loss relationship.
For simplicity, we will assume throughout that $C$ is
sufficiently small that we can neglect any small size
cutoff.

The PLR model is defined in terms of noncumulative
probabilities $p_i$, but to reliably compare with data it
is necessary to use cumulative distributions. Since ${p}_i
\propto (l_i)^{-(1+\alpha)}$ (Eq.~(\ref{pdist})), the naive
expectation is that the cumulative distribution $P(\geq
l_i)\propto (l_i)^{-\alpha}$. However, this is not
necessarily the case for discrete data sets, where
cumulative distributions are attained by summing, rather
than integrating the density. In fact, in the discrete
case, the cumulative distribution can be steeper,
shallower, or have the same decay properties as the
density, depending on how densely the data is sampled.
Thus, unlike the case of a  continuous probability density,
there is no general relationship between discrete
probability distributions and their noncumulative
densities.  We cannot simply assume that since ${p}_i
(l_i)$  is a power law with slope ${-(1+\alpha)}$, that
$P( \geq l_i)$ is a power law, let alone with slope
${-\alpha}$. This issue is fundamental in the theory of
discrete probability distributions, and also arises for the cuts model
(Section IV), which is also inherently discrete), and in
comparing PLR with the continuum and cuts models (Section
V).

Furthermore, in making comparisons with data, use of the
density, rather than the distribution does not solve the
problem. Use of the cumulative distribution is in fact
preferable, because it avoids statistical anomalies
associated with binning. The cumulative distribution simply
corresponds to a normalized plot of the ranked (by size)
order of events in a catalog, which does not introduce any
statistical biases. 

Although the PLR model can be used to generate a
cumulative event probability function $P(\geq l)$ which is
inherently discrete, most data sets exhibiting power laws
in the cumulative event  probability as a function of size
are sufficiently dense to exhibit a fairly convincing unit
difference in slope between the density and the cumulative
distribution. This leads us to determine circumstances
under which the naive expectation of unit difference in the
exponent between the cumulative distribution and the
noncumulative density is in fact correct.

This requires sampling in the data set which is sufficiently dense
that integration of the density to obtain the cumulative distribution
is a good approximation to computing the discrete sum.
One possible explanation is to hypothesize
that most data sets are {\em mixtures} from many different
systems, or the same system averaged over long times. Thus a
complete treatment of how to assess whether data is
consistent with a PLR mechanism ultimately requires a
treatment of mixtures \cite{FiniteMixtures, mixtures}.

The simplest scenario corresponds to a mixture of  discrete power law
distributions with the same exponent. This  generates a power law
with that same exponent, but possibly different short- and
large-scale cutoffs, and provides
a simple and unambiguous way to connect the PLR ${p}_i
\propto (l_i)^{-(1+\alpha)}$ with $P(\geq
l_i)\propto (l_i)^{-\alpha}$.
This scenario assumes  sufficient data up to some cutoff
size $L$,  binned  with fixed $\Delta l$, to treat
the resulting ${p}_i$ as binned samples from a
continuous density. Then we can define
\begin{equation}
\label{cum2}
\begin{array}{l}
 P( \geq l_i ) = \sum\limits_{j \geq i} {\left( {l_j  + C }
 \right)^{ - \alpha - 1} } \left( {l_{j + 1}  - l_j }
 \right) \\
 =\sum\limits_{j \geq i} {\left( {l_j  + C }
 \right)^{ - \alpha - 1} } \Delta l \\
\end{array}
\end{equation}
which in the limit of large data sets approximates a
continuous $P(\geq l)$ satisfying
\begin{equation}
\label{cum3}
\begin{array}{l}
  P(  \geq l) \propto \int\limits_l^L {p({\bf x})dx = }
  \int\limits_l^L {\left( {{\bf x} + C }
  \right)^{ - \alpha - 1} d{\bf x}} \\
  \propto \left( {\left( {l + C } \right)^{ -
  \alpha}  - \left( {L + C } \right)^{ -
  \alpha} } \right)\quad \quad  \\
 \end{array}
\end{equation}
leading to the exponent $\alpha=1$ in $d=1$.
Table~\ref{tbl:compare} assumes these properties of the PRL
model. Note, however, that when the PRL model is used  and
the $l_i$ are not densely sampled, then the above
calculations for the cumulative distribution need not hold.


\section{IV. Cuts Model}

The cuts model~\cite{PRE1, INFOCOM} is a simple, analytic
model that helps clarify the discrepancy between the power law
exponents predicted by the continuum and PLR models. We
focus on $d=1$ for this case and the comparisons. Higher
dimensional generalizations of the cuts model are possible,
but correspond to constrained optimization schemes (e.g.
the grid design problem in \cite{PRE1}) or choices of
$p({\bf x})$ with special symmetries. As we show below, the
other two models as formulated above agree with the cuts
model in (different) asymptotic regimes. We also use the cuts model to
formulate an extension of the PLR model that describes the dense
resource limit, where all three models agree.

Like the continuum model, the cuts model  is naturally
understood as a continuum limit of a percolation lattice
model, but it is a variant of percolation  which  includes
an explicit constraint on the resources, as in PLR.
The cuts model removes the assumption that
the event sizes $l(x)$ are nearly continuous (an
approximation made in the continuum model), which makes it possible to
span both the dense and sparse resource regimes in a single
formulation of the model.

Consider a percolation forest fire lattice model in one
dimension. Resources are vacancies that act as dividers or
{\em cuts} between connected clusters of occupied sites. An
example of this is shown for $d=1$ in
Figure~\ref{1d_lattice}. If we  take a continuum limit by
rescaling into a finite interval and taking the number of
lattice sites to infinity, the the cuts become
infinitesimally thin, zero dimensional dividers between
continuous connected regions of unit density.

\begin{figure}
\resizebox{!}{.22\textwidth}{{\includegraphics{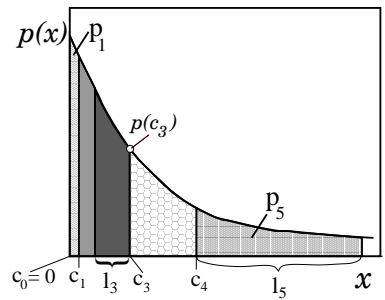}}}
\caption{\label{cuts_nc} Illustration of cuts model mapping from
probability function $p(x)$ which is a continuous function of the
spatial coordinate $x$ to a discrete set of probabilities $p_i$.
The
cut positions chosen to optimize a yield function, $Y$ or
$Y^{t}$.}
\end{figure}

The cuts model is defined on position space $x$, $x \in [0,
\Lambda] \subset \Re$, where $\Lambda$ is the large-scale
cutoff. A discrete set of zero-dimensional cuts divide the
axis into  a set of separate one-dimensional line segments.
The model imposes the constraint that the  maximum number
of cuts is a natural number $N$. Analogous to the PLR
model's explicit constraint on total resources ($\sum r_i
\le R$), optimal solutions make full use of all available
resources ($\sum r_i = R$ in PLR and $\# $ cuts=$N$ in
cuts). Events are triggered (sparked) according to a
spatial probability function $p(x)$ as in the continuum
model, propagating along the connected cluster, between
adjacent cuts. The position of the $i$th cut is labeled
$c_i$ and $c_0$ is at $x = 0$. The cut positions define
discrete line segments $l_i$ and the corresponding event
probabilities $p_i$:
\begin{eqnarray}
\label{cuts_def}
  l_i &\equiv& c_i - c_{i-1} \nonumber \\ {\rm and}\ \ p_i
&\equiv&
  \int_{c_{i-1}}^{c_{i}} p(x) dx.
\end{eqnarray}
In other words, the cuts map the continuous spatial
function $p(x)$ defined on $[0, \Lambda] \subset \Re $ into a
discrete set of events with probability $p_i$ given by the
cumulative probability of sparking the segment of length
$l_i$ between adjacent cuts.  This mapping is illustrated
in Figure~\ref{cuts_nc}.

Carlson and Doyle~\cite{PRE1} maximized the yield function
\begin{equation}
\label{cuts_yield}
Y = 1-c \sum p_i l_i
\end{equation}
 with respect to the cut positions. Note that this is the same
yield
function used in the PLR model. They found an iterative solution
for
the optimal cut positions in the continuum limit:
\begin{equation}
\label{cuts_recur1}
  p_i + l_i\: p(c_i) = p_{i+1} - l_{i+1}p(c_i).
\end{equation}
Unfortunately, analytic solutions to this equation involve
transcendental functions even if there is a simple
functional form for $p(x)$.

The problem simplifies if we consider a slightly modified
cost function, replacing $J= \sum p_{i} l_i$ in
(\ref{cuts_yield}), with $J^t= \sum p_{i}^{t} l_i$,  where
$p_{i}^{t}$ is the probability of events of index greater
than $i$, $p_{i}^{t} \equiv \sum_{j=i}^{N} p_j$. This cost
function can be naturally motivated in many cases, such as
web layout~\cite{INFOCOM}. Furthermore, as we show below,
results obtained for the power laws using this modified
cost function are equivalent to the original cost function
in the small and large size asymptotic regimes.

With the modified cost function, we can define the yield as
\begin{equation}
\label{cuts_yield2}
 Y^t = 1 - c \sum p^{t}_{i} l_i,
\end{equation}
and optimize the yield with respect to the cut positions
${c_i}$ by setting $\partial Y^{t} / \partial c_i = 0$.
Using the definitions from Eq.~(\ref{cuts_def}), the
following iterative equations hold for the optimal cut
positions:
\begin{equation}
\label{cuts_recur2}
  p_i = l_{i+1} \: p(c_i).
\end{equation}
This equation is simpler to iterate than
Eq.~(\ref{cuts_recur1}). Its solutions are no longer
transcendental functions, and optimal $l_i$ for general
$p(x)$ can easily be found using simple numerical
techniques. Note that the number of cuts $N$ does not
appear explicitly in the recursion relation. Instead, the
equation requires two initial cut positions, $c_i$ and
$c_{i-1}$(which is the lower limit of integration for the
integral defined as $p_i$). These initial cut positions
define a length scale, $l_i = c_i - c_{i-1}$. This length
scale together with the large-scale cutoff, $\Lambda$
determine the total number of cuts, $N$. Therefore,
choosing two initial cut positions is equivalent to
specifying $N$ for a fixed $\Lambda$.

\subsection{Cuts model for an exponential  distribution of sparks}

\begin{figure}
\resizebox{!}{.22\textwidth}{{\includegraphics{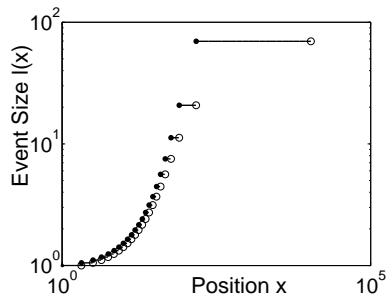}}
} \caption{\label{lx_figure} Event size $l(x)$ as a
function of $x$ for the cuts model. When the value of
$l(x)$ is small (and $x$ is small) the function is nearly
continuous but when the value of $l(x)$ is large the
function is piecewise constant, displaying obvious discontinuities.}
\end{figure}

To solve the recursion equation analytically we first
choose $p(x) = \lambda e^{-\lambda x}$, which leads to an
especially simple solution:
\begin{equation}
\label{cuts_2}
p_{i}^{t} = e^{ -\lambda c_{i-1}}.
\end{equation}
As with the other two models, we are interested in the
probability distribution of event sizes $\rho(l_i)$ and the
cumulative probability distribution $P( \geq l_i)$. In this
case, solving for  $P( \geq l_i)$ is transparent;
\begin{equation}
\label{cuts_3}
P(\geq l_i) = p_{i}^{t}.
\end{equation}
We substitute $p(x)$ into Eq.~(\ref{cuts_recur2}) to find a
recursion relation for the optimal region sizes:
\begin{equation}
\label{l_iter}
l_{i+1} = \frac{e^{\lambda l_i} -1}{\lambda}.
\end{equation}
Notice that the event sizes increase exponentially as $l_i$
becomes large. We use $l_i$ to construct the function $l(x)$
which is defined as the event size $l_i$ when a spark hits
site  $x$. This function is piecewise constant between
cuts, as illustrated in Figure~\ref{lx_figure}.  For large
$x$, the function exhibits large discontinuities.  For
small $x$, while still discrete, it approaches a continuous
function.

The slope of $P( \geq l_i)$ on a log-log plot can be
easily calculated in limiting cases by substituting
Eq.~(\ref{l_iter}) into Eq.~(\ref{cuts_2}), dividing
$\Delta \log p^{t}_{i}$ by $\Delta \log l_{i}$, Taylor
expanding, and dropping higher order terms. The limiting
case describing the large event sizes, with sparse resource
allocations, is discussed in~\cite{INFOCOM}. Following the
derivation there:

\begin{eqnarray}
\label{cuts_limit1}
\lim_{l_i \rightarrow \infty}  \frac{ \log p_{i+1}^{t}-\log
p_{i}^{t}}{\log l_{i+1}- \log l_{i} } &= \lim_{l_i \rightarrow
\infty}
\frac{\log e^{- \lambda c_i} - \log e^{- \lambda c_{i-1}}}{\log
\frac{
e^{\lambda l_i} -1} {\lambda l_{i}} } \nonumber\\ =\lim_{l_i
\rightarrow \infty}& \frac{- \lambda l_i}{\log ( e^{\lambda l_i}
-1) -
\log \lambda l_{i}} =  -1
\end{eqnarray}
The opposite limiting case, describing small events, and
high resource densities, can also be calculated. We find:
\begin{eqnarray}
\label{cuts_limit2}
\lim_{l_i \rightarrow 0}  \frac{ \log p_{i+1}^{t}-\log
p_{i}^{t}}{\log
l_{i+1}- \log l_{i} } &=&\lim_{l_i \rightarrow 0} \frac{- \lambda
l_i}{\log \frac{ e^{\lambda l_i} -1} {\lambda l_{i}}  }\nonumber
\\
=\lim_{l_i \rightarrow 0}  \frac{- \lambda l_i}{\log ( 1 +
\frac{(\lambda l_i)^2}{2 \lambda l_i}) } &=& -2
\end{eqnarray}

\begin{figure}
\resizebox{!}{.25\textwidth}{{\includegraphics{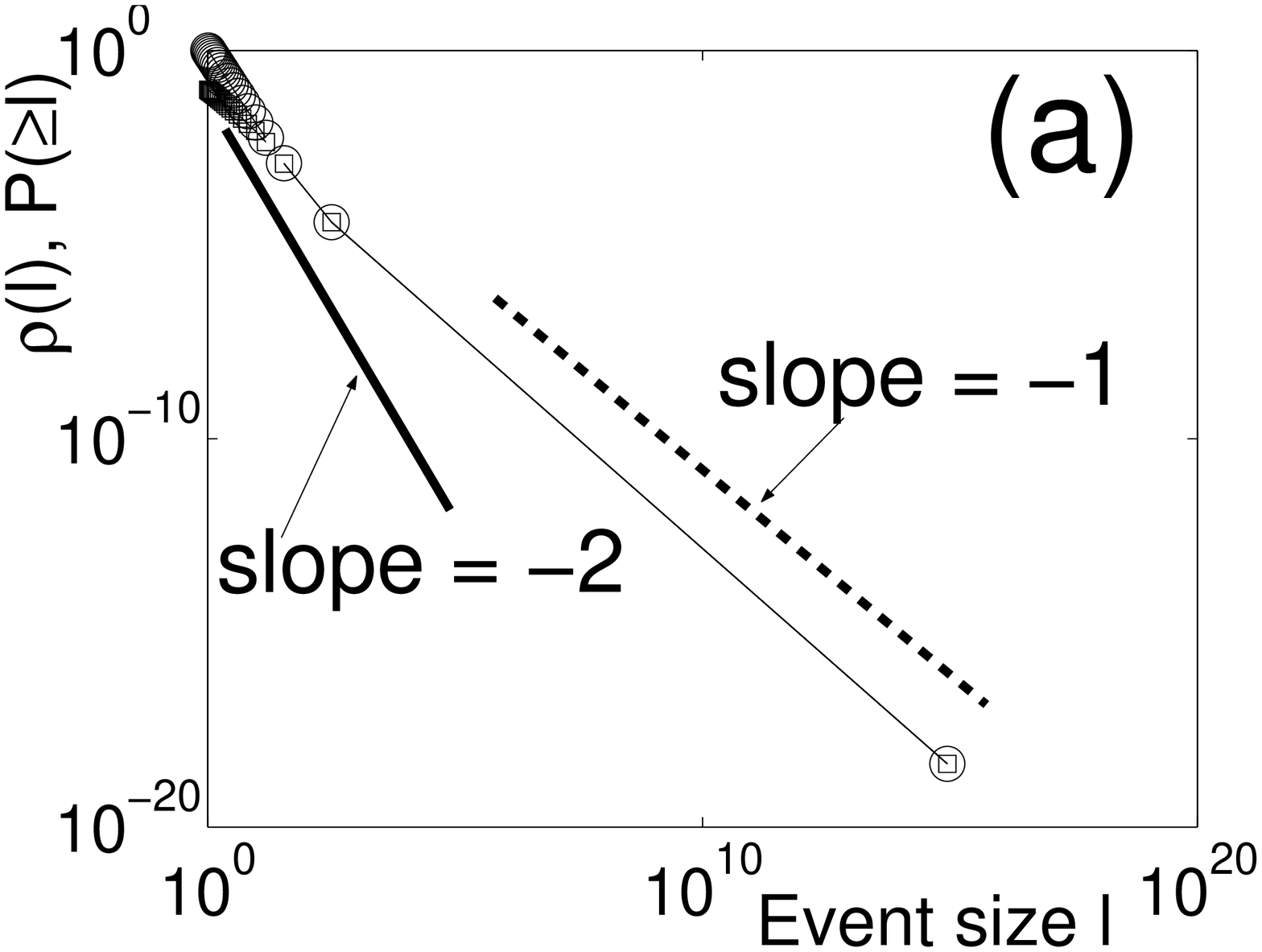}
}}
\resizebox{!}{.25\textwidth}{{\includegraphics{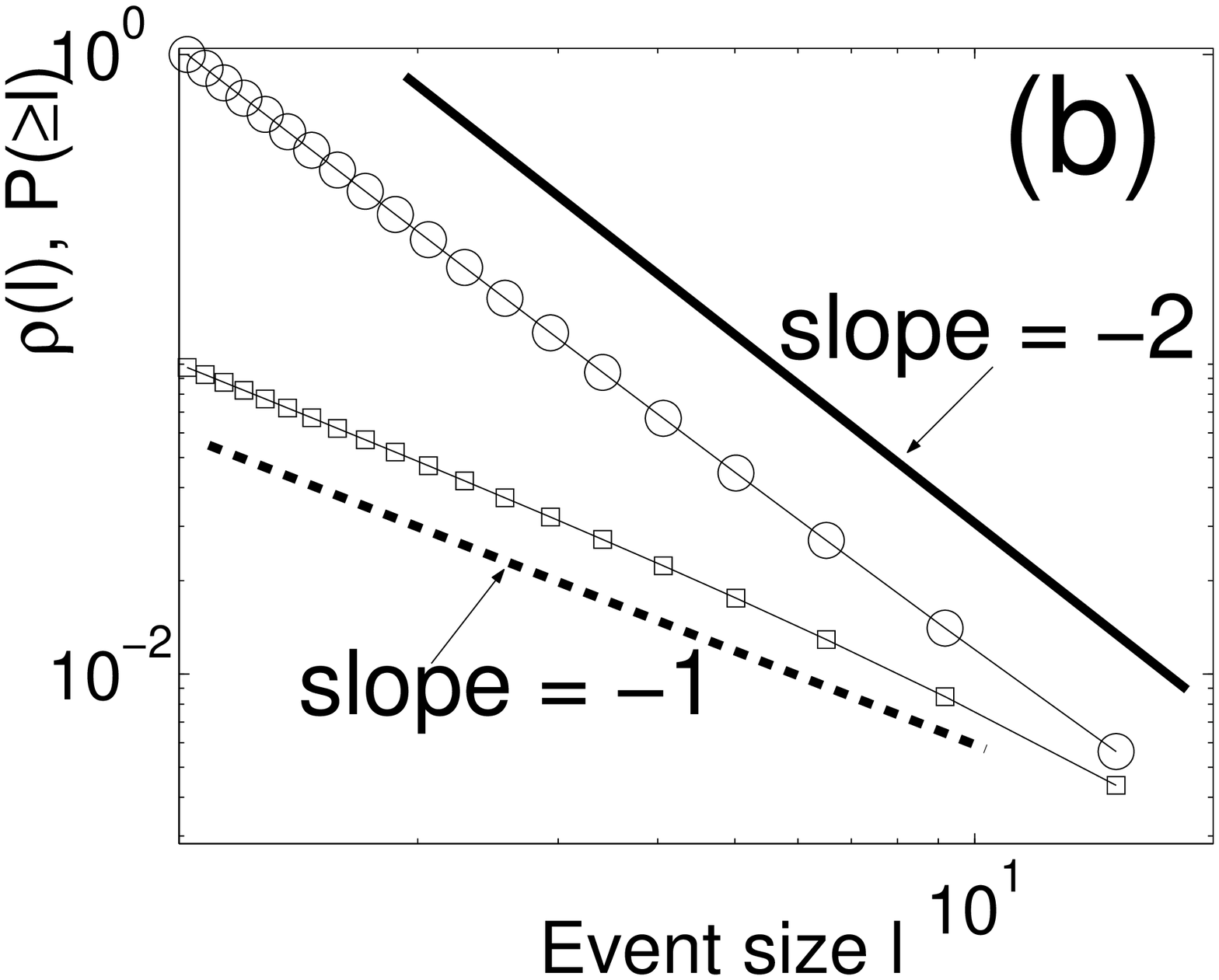}
}}
\caption{\label{cuts_figure}
Numerical results for the cuts model with an exponential distribution of
sparks. Figure (a) shows the cumulative probability
$P(\geq l)$ (Large circles) and probability density $\rho(l)$ (small squares) vs. 
event size $l$ for  the one dimensional exponential 
cuts model (i.e.~$p(x)=\lambda e^{-\lambda x})$ with 
the modified yield function (using ${p^t}_i$). Points are calculated from
Eq.~(\ref{cuts_recur2}) iterated backwards and forwards from the initial 
cut positions $c_i = 300, c_{i-1} = 250$.  The 4th iteration forwards 
results in  a data point too large to compute. The solid line illustrates 
a power law with exponent $-2$, and the dashed line illustrates exponent 
$-1$. Figure  (b) is an enlarged view of (a) in the region where $l$ is 
small. For small $l$, the cumulative probability has a steeper slope than 
the probability density, but for large $l$ their slopes are the same. 
Again, the solid line illustrates exponent $-2$ and the dashed line is $-1$. } 
\end{figure}

\begin{figure}
\resizebox{!}{.25\textwidth}{{\includegraphics{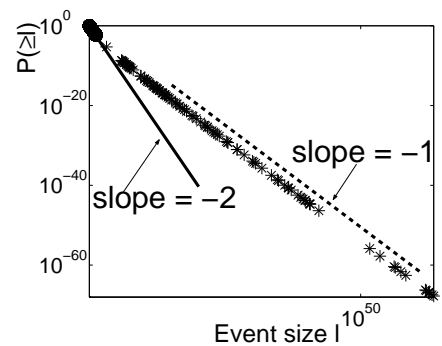}
}}
\caption{\label{cuts_mixture}
Cumulative probability $P(\geq l)$ vs. $l$ for a mixture of data sets with 
slightly different large scale cutoffs, $\Lambda$. These data sets are 
generated  as in Figure \ref{cuts_figure}, but sampling the initial, seed cut 
positions randomly, and combining data fro the different choices. The second 
cut position $c_i$ was chosen from a uniform 
distribution on $[200,300]$ and the distance between the first and second 
cut $l_k$ was chosen from a uniform distribution on $[25,50]$.  
Note: $c_{i-1} = c_i - l_i$.} \end{figure}

We can also investigate the asymptotic behavior for small
and large events in this model numerically by choosing two
initial cut positions $c_i$ and $c_{i-1}$ and then
iterating Eq.~(\ref{cuts_recur2}) backwards and forwards.
The cumulative probability $P( \geq l_i)$ (large circles) vs.
the event size $l_i$ is shown in
Figure~\ref{cuts_figure}(a) and (b), and the limiting power
law behaviors in the small and large event size limit
derived analytically are apparent. Notice that there are
only a few points in the slope = -1 regime in
Figure~\ref{cuts_figure}(a).  This is because the event
sizes are increasing exponentially as shown in
Eq.~(\ref{l_iter}).  We can populate the tail of this
distribution by combining many data sets with slightly
different initial cut positions, $c_i$, $c_{i-1}$, and the
results are shown in Figure~\ref{cuts_mixture}. This
models a mixture of data from systems with the same number
of resources $N$ but different large scale cutoffs,
$\Lambda$.

Figure~\ref{cuts_figure}(a) and (b) also show the
probability density $\rho(l_i)$ (small squares) vs. the
event size $l_i$. For small $l$, the cumulative probability
has a steeper slope than the probability density, while for
large $l$ the points are very nearly the same and have the
same slope.  This occurs because the probability density
for the cuts model is inherently discrete.  As discussed
in Section III, if
a probability density is smooth and continuous, the
corresponding cumulative probability can be found by
integrating the density. For example, if $p(x)$ is a power
law with exponent $-(\alpha+1)$, then the cumulative
probability is a power law with exponent $-\alpha$, as
we intuitively expect.  This simple, intuitive result also applies when
data  consists of a set of discrete probabilities $p_i$ which are
sufficiently dense that we can use them to derive a continuous
probability distribution as we did in the PLR model
Eq.~(\ref{cum2}).  However, in Figure~\ref{cuts_figure} the
discrete probabilities {$p_i$} are not dense, and the
relationship between $P(\geq l)$ and $p_i$ is not the same as
in the continuous case. As $l_i$ becomes large,
Eq.~(\ref{l_iter}) indicates $l_{i+1}>> l_i$, and $p_{i+1} <<
p_i$. The cumulative probability distribution becomes the same
as the probability density in the tail:
\begin{equation}
P(\geq l_i) = \sum_{j=i}^{\infty} p_j \simeq p_i
\end{equation}
This asymptotic behavior is verified in
Figure~\ref{cuts_figure}(a) and Figure~\ref{cuts_mixture}.

\subsection{Cuts model for a  power law distribution of sparks}

We can also analytically solve for the optimal cut positions in
the case of a power law distribution of sparks:
$p(x) = a x^{-(a+1)}$ \cite{a_2}.
Using the same procedure as in
the exponential case, we find the following results for
the discrete probabilities and the corresponding event
sizes:
\begin{eqnarray}
p_i &=& (c_{i-1})^{-a} - (c_{i})^{-a} \nonumber \\ l_i &=&
\frac{p_i}{p(c_i)}
\label{cuts_power_eq}
\end{eqnarray}
so that
\begin{equation}
l_i = \frac{(c_{i-1})^{-a} -
(c_{i})^{-a}}{a(c_i)^{-(a+1)}}
\label{cuts_recur_power}
\end{equation}
For $a >1$,  the slope of $P(\geq l)$ vs. $l$
on a log-log plot approaches $-2a/(a+1)$ as $l$ becomes
small.  As $l$ becomes large, the slope approaches $-1$.
These asymptotic relationships are derived
in Appendix A.  In addition, as $a$
approaches infinity the initial probability density $p(x) =
a x^{-(a+1)}$ decays faster than any power law. 
Notice that in the limit $a
\rightarrow \infty$ we recover $-2$ as the exponent for the
cumulative probability distribution, which is exactly the
same as the exponential result.

\begin{figure}
\resizebox{!}{.25\textwidth}{{\includegraphics{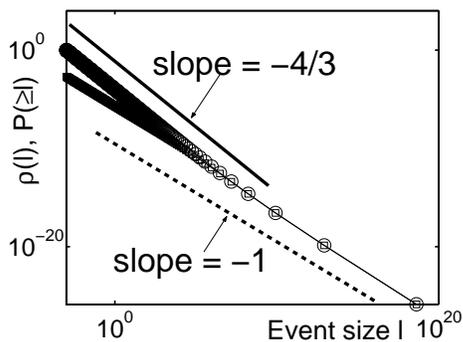}}}
\caption{\label{cuts_power} Cumulative probability $P(\geq l)$ (Large
circles) and
probability density $\rho(l)$ (small squares) vs. event size $l$ for the cuts model 
with  power law spark  probability density with parameter $a=2$ (in $p(x)=ax^{-(a+1)}$). 
Points are calculated from Eq.~(\ref{cuts_recur_power}) iterated forwards 2000 
times from the initial cut positions $c_{i-1}=1, c_{i}=1.001$.  The solid line 
illustrates a power law with exponent $-4/3$, and the dashed line illustrates 
exponent $-1$.} \end{figure}

We also investigate the event size distribution for power
law $p(x)$ by solving the recursion relation in
Eq.~(\ref{cuts_recur_power}) numerically.
Figure~\ref{cuts_power} shows the cumulative (large
circles) and noncumulative (small squares) event size
distributions for a  power law spark distribution with
$a=2$ (i.e. $p(x)=a x^{-(a+1)}=2 x^{-3}$). For small $l$
this leads to a cumulative probability distributions of
event sizes that has a shallower slope than  the corresponding 
data for the exponential spark density (Figure \ref{cuts_figure})
but still a steeper slope than for larger
events. The slope of the cumulative distribution  is
close to the analytically calculated asymptotic value of
$\alpha=-2a/(a+1)= -4/3$ in the small event limit (Appendix
A). For large $l$ the slope is approximately $-1$.
The corresponding data for the case $a\to 1$ ($p(x)\sim 1/x$)
has slope $-1$ for the entire range of event sizes.
Additionally,  solutions of the cuts model obtained for
a power law distribution of sparks has the feature that the
large event sizes $l_i$ increase at a slower rate than in
the corresponding exponential solution. Therefore we are
able to see more points in the tail of
Figure~\ref{cuts_power} and easily confirm the slope
$-1$ that we derive analytically (Appendix A).

\section{ V. Comparing models}

We  next  make  more direct comparisons between
the continuum, PLR, and cuts models. Despite the apparent
differences, we show that there are a variety of
cases where  one model can be used to
approximate another. In these cases the resulting power
laws match.  However, in doing this we face several challenges:
\begin{itemize}
\item The PLR and continuum models use the expected event
size as the cost function: $J =\sum p_i l_i$ (yield
function $Y$ in Eq.~(\ref{cuts_yield})). The cuts
model is most easily solved analytically
for  the cumulative cost function $J^{t} = \sum
p^{t}_{i}l_i$ (yield function $Y^t$ in
Eq.~(\ref{cuts_yield2})), where ${p^t}_i=\sum_{i\le j\le N}
p_j$.
\item The continuum and cuts models specify a probability
density $p(x)$ which is a continuous function of the
spatial position  $x$, while the PLR model specifies
condition categories $i$ with discrete probabilities $p_i$
which have no {\it a priori}  association with a position $x$.
\item The cuts and PLR models specify a set of discrete
probabilities $p_i$ and corresponding set of discrete event
sizes $l_i$, while the continuum model uses only continuous
$p(x)$ and $l(x)$.
\item The cumulative distribution $P(\geq l)$ is an
analytic function of the probability density $\rho(l)$ only
if the density is a {\em continuous} function of event size.
If instead the density $\rho(l)$ (or $p_i(l_i)$)
is discrete, as it is for
the cuts and PLR models, there is no universal
analytic relationship
between the cumulative and noncumulative distributions.
\end{itemize}
We address all  these issues in the subsections that follow.

\subsection{Comparing results obtained for different cost functions}

To reconcile the cost functions of the different models we
can either  find solutions to a ``$J$-cuts model'' which
uses the original cost function $J$, or we can  adapt the
PLR model to use the modified cost function $J^{t}$. The
recursion relation for the cuts model with the cost
function $J$, Eq.~(\ref{cuts_recur1}) is more difficult to
solve, but fortunately we can determine the asymptotic
behavior of this ``$J$-cuts model'' without solving those
equations. This is because the asymptotic results in the
simple exponential ``$J^{t}$-cuts model''
(Eqs.~(\ref{cuts_limit1}) and (\ref{cuts_limit2})) are
valid for both cost functions $J$ and $J^t$. In particular,
the optimal solutions $\{l_i\}$ are asymptotically equal
for the two costs ($J^{t}$, $J$) in the limits $l_i
\rightarrow \infty$ and $l_i \rightarrow 0$. Proof of this
result is given in Appendix B. This implies that our
results for the cuts model can be directly compared to the
results for the PLR  and continuum models in these limiting
cases, as shown in Table I. Alternatively, we can modify
the PLR model to use the same cost function as the
``$J^{t}$-cuts model''. This is particularly simple if the
probability distribution of sparks $p(x)$ is
exponential, since cumulative and noncumulative exponential
distributions are proportional to each other. 

\begin{figure}
\resizebox{!}{.22\textwidth}{{\includegraphics{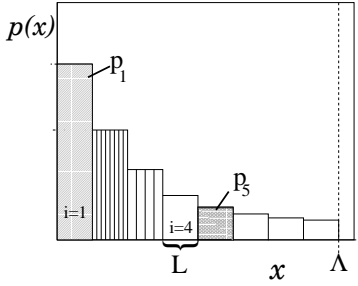} }}
\caption{\label{PLR_of_x} We use the category size $L$ to
generate a piecewise constant function $p(x)$ of position $x$
from the discrete set of probabilities $\{ p_i\}$ in PLR.}
\end{figure}

\subsection{Mapping of the PLR event categories to spatial positions}

To compare PLR to other models, we also must decide how to
associate the discrete probabilities $p_i$ in the PLR model
with positions $x$. In the PLR model, we derive scaling
relations between resources and event sizes by imagining
that each event category $i$ is associated with a region of
the {\em same} total length $L$, inside of which the
probability is a uniform $p_i$ as illustrated in Figure
\ref{PLR_nc}. (The length $L$ is later divided up into
optimal event sizes $l_i$.) This procedure is discussed in
Section III.

To construct a mapping  from the event categories to the
real axis, we  can  use this length $L$ to derive a
right-continuous piecewise constant probability function
$p(x)$ on the real line, as illustrated in Figure
\ref{PLR_of_x}. We order the $p_i$ so that they are
monotonically nonincreasing, associate each category $i$
with a length $L$, and place the categories adjacent to one
another on the real line. Then $p(x) = p_i$ whenever $x \:
\in \: [(i-1)L, i L)$. We can then use PLR formalism
to calculate the optimal event sizes $l_i$ within each
category. This defines a event size function $l(x)$ which
describes the size of the loss which occurs when a spark
hits position $x$. We define $l(x) = l_i$ whenever $x\:
\in \: [(i-1)L, i L)$. Note that $L$ is the maximum
possible event size in PLR and the large-scale cutoff
$\Lambda$ is defined by $L$ and the number of event
categories $n$: $\Lambda = n L$.

Intuitively, here it is helpful to think of the PLR model
as a coarse-grained version of the cuts model. The
piecewise constant spark  probability density $p^{PLR}(x)$
can be viewed as an approximation to some underlying
continuous probability density  $p^{cuts}(x)$ which has
been averaged to produce a constant value over  each
interval of length $L$. As $L$ becomes smaller, PLR becomes
a better approximation to the cuts model with a continuous
$p(x)$.

\subsection{Comparing PLR and cuts}

\begin{figure}
\resizebox{!}{.25\textwidth}{{\includegraphics{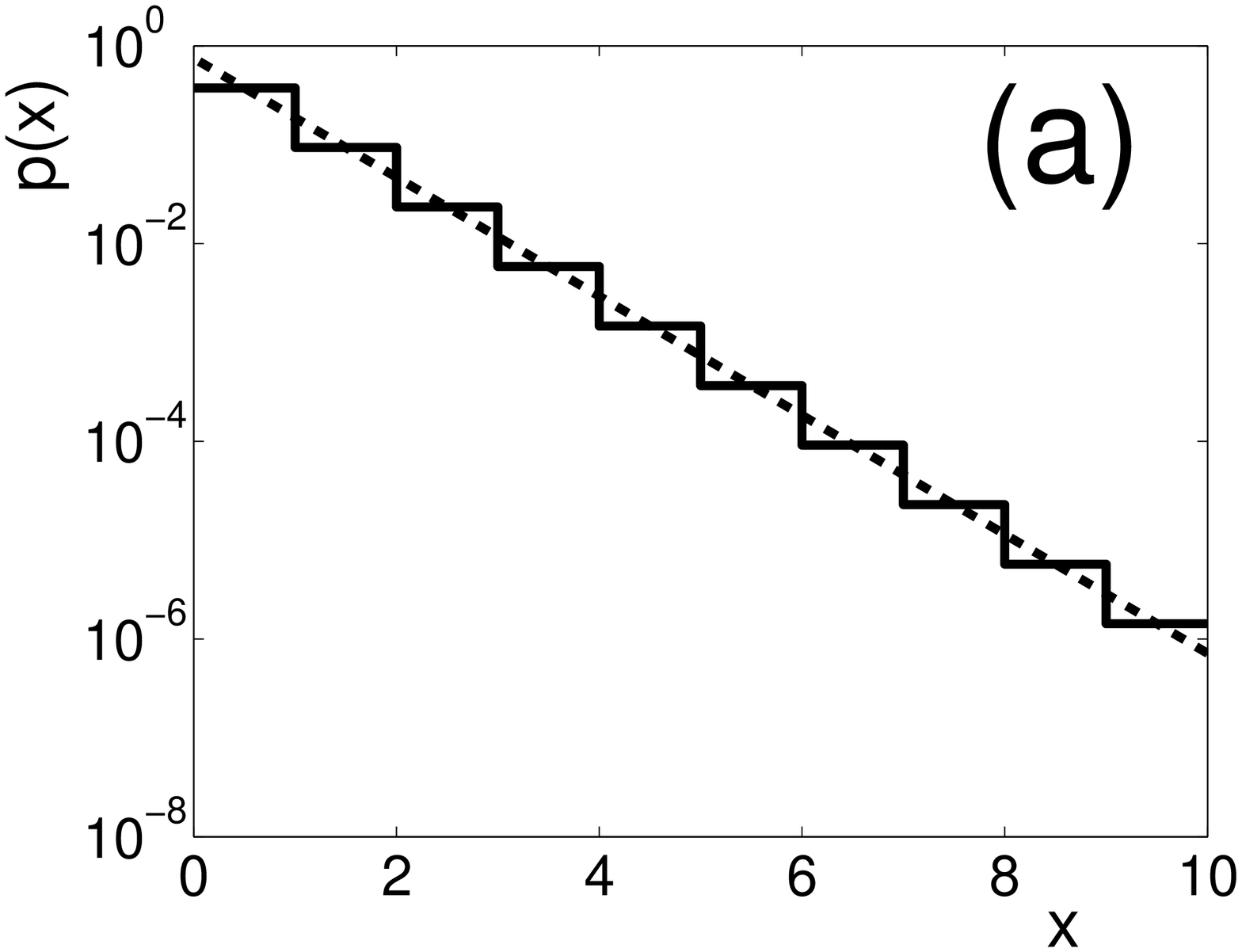}
}}
\resizebox{!}{.25\textwidth}{{\includegraphics{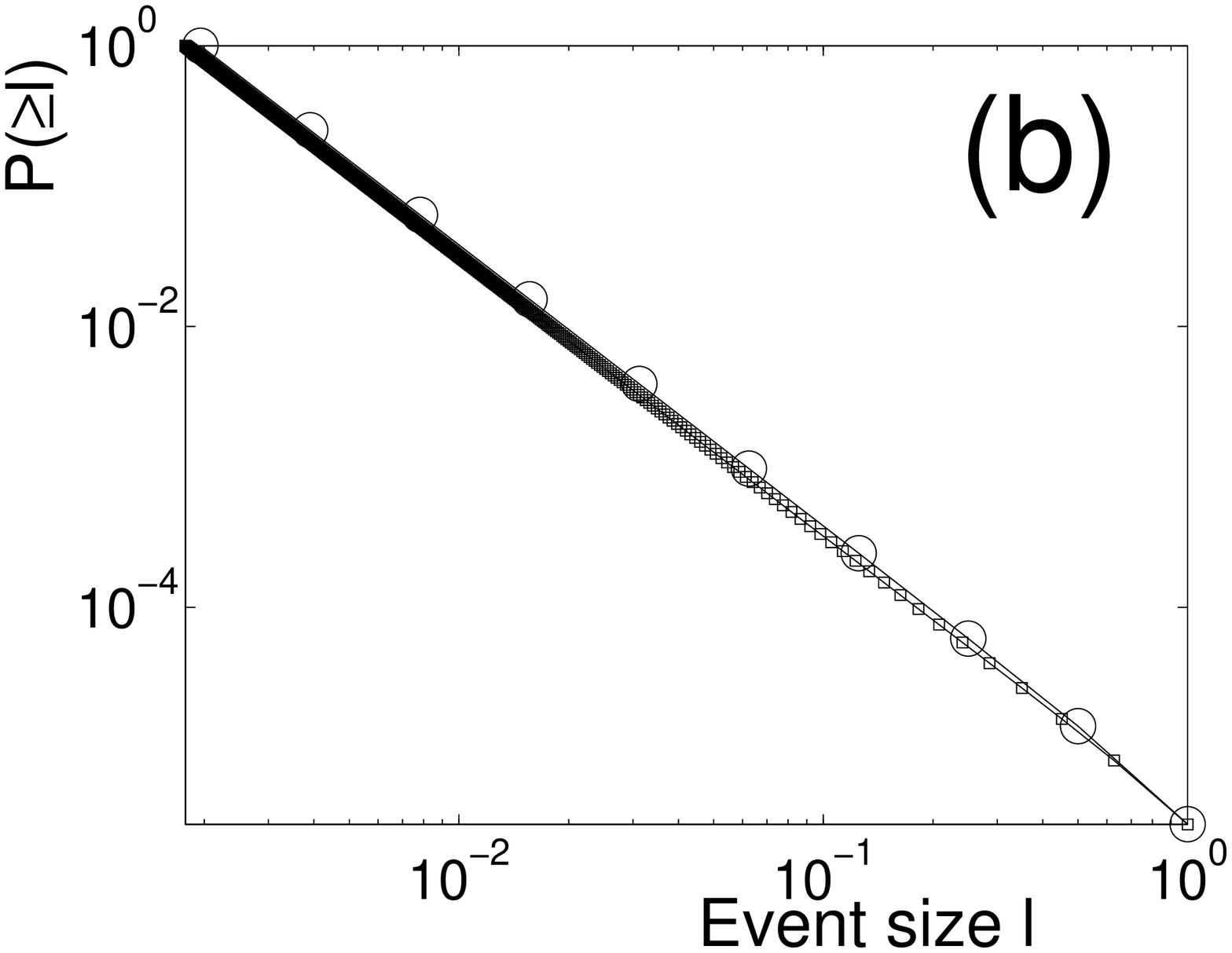}
}} \caption{\label{comp} Comparing the PLR and cuts models.
(a) Spark probability density
functions on a semilog plot. The solid line represents the
piecewise constant function $p^{PLR}(x) = K_1 e^{-\lambda
(i-1/2) L}, x \: \epsilon [(i-1)L, iL), L=1, \lambda =
\log(4)$, and the dashed line represents the continuous
probability function $p^{cuts}(x) = K_2 e^{-\lambda
x},\lambda = \log(4)$. Constants $K_1$ and $K_2$ are chosen
so the probability densities are normalized on $x \epsilon
[0, \Lambda=10]$. (b) Cumulative probability $P(\geq l_i)$
vs. event size $l_i$ for the PLR model (large circles) and
the cuts model (small squares).}
\end{figure}

The cuts and PLR models can compared in many regimes
because they both produce inherently discrete event size
distributions. The spatial mapping of event categories to
spatial positions, and the approximation of a continuous
$p(x) =p^{cuts} (x)$ (for cuts) by a piecewise continuous
$p^{PLR}(x)$ composed from the $p_i$'s lead to excellent
agreement between the two models for a wide range of
$p(x)$. For the cuts model we choose a continuous
probability density:
\begin{equation}
p^{cuts}(x) = K_2 e^{-\lambda x}
\end{equation}
For the PLR model we choose a probability density which is
piecewise constant on intervals of length $L$:
\begin{equation}
p^{PLR}(x) = K_1 e^{-\lambda (i-(1/2)) L},\: \: x \:
\epsilon \: [(i\!-\!1)L,\; iL)
\end{equation}
and chose these densities so that $p^{cuts}(x)$ matches
$p^{PLR}(x)$ at the mid-point of each interval. The density
$p^{PLR} (x)$ can be thought of as a coarse-grained average
of $p^{cuts} (x)$. Graphs of these functions are shown in
Figure~\ref{comp}(a). We can trivially modify the PLR model
to use the same cost function $J^{t}$ because cumulative
and noncumulative exponential distributions are
proportional to each other, implying $p^{t}_i \propto p_i$.

Next we  use the PLR model to find the optimal $l_i$, and
thus $p_i(l_i)$ and $P(\geq l_i)$ for the  spark
probability density $p^{PLR}(x)$. We take  $L=1$ and a
large scale cutoff $\Lambda$ which is $n=10$ times larger
than $L$. The cumulative probability $P(\geq l_i)$ vs.
event size $l_i$ (large circles) is shown in
Figure~\ref{comp}(b). Note that $P(\geq l_i)$ has a
exponent of $-2$, which is exactly the same as the exponent
for the noncumulative probability $p_i(l_i)$. Again, this
is due to the discrete nature of $p_i(l_i)$ and the
exponential $p(x)$ which is approximated by the piecewise
constant $p^{PLR}(x)$.

To obtain the corresponding solution for the event size
distribution of the cuts model, we use the recursion
relation (Eq.~(\ref{cuts_recur2})) to compute the  optimal
$l_i$, $\rho(l_i)$ and $P(\geq l_i)$ for the continuous,
exponential $p(x)=p^{cuts}(x)$. We choose the initial cut
positions based on our solution for the largest event
obtained for the corresponding PLR model above. In other
words, we take $c_i = \Lambda$ (the endpoint of the
interval on the real axis for the mapping of the PLR
categories into position space) and $c_{i-1} = \Lambda -
l_n$, where $l_n$ is the largest event size in the PLR
model. We then iterate the recursion relation
Eq.~(\ref{cuts_recur2}) backwards until we reach the cut at
position $x=0$. The cumulative probability $P(\geq l_i)$
vs. event size $l_i$ (small squares) is shown in
Figure~\ref{comp}(b). $P(\geq l_i)$ has a exponent of $-2$
in agreement with PLR for the same cost function $J^t$, and
the corresponding spark distributions $p^{cuts}(x)$ and
$p^{PLR}(x)$.

Thus the cumulative probabilities for the cuts and PLR
models are remarkably similar. This indicates that even
outside the asymptotic regime ($l_i \rightarrow 0$),
the cuts model and the PLR model match for an exponential
spark probability density. Note that in this example we
are still in the regime where the cuts model solution
$P(\geq l_i)$ vs. $l_i$ has a slope of $-2$ on a log-log
plot --- that is, the dense resource regime.

\subsection{Connections between the continuum model and
the discrete PLR and cuts models}

We next compare the continuum model, which has a continuous
event size function $l(x)$, with the cuts and PLR models
which both have a piecewise constant $l(x)$, corresponding
to the discrete $l_i$ for these models (and the spatial
mapping, in the case of PLR). The continuum model cannot be
extended outside of the dense resource regime, because it
builds in the assumption of a continuous event size
function $l(x)$. Interestingly, all three models can be
made to agree  in the dense resource limit. For the PLR and
cuts models, these correspond to regimes in which the
piecewise constant function $l(x)$ becomes nearly
continuous. We begin by comparing the continuum model to
the cuts model. The cuts model predicts that for small
event sizes (and thus dense resource allocations), the
function $l(x)$ will be close to continuous (as shown in
Figure~\ref{lx_figure}). We  showed earlier in
Eq.~(\ref{cuts_limit2}) that in the limit $l_i \rightarrow
0$, the $J^{t}$-cuts model predicts a power law with
exponent $-2$. In Appendix B, we show that the solution for
the cuts model with the modified cost function $J^t$ is the
same as the solution for the cuts model with the original
cost function $J$ in this limit. Therefore that the cuts
model matches the continuum model in the limit $l_i
\rightarrow 0$,  when the two models have the same cost
function $J$. Note that even though $l(x)$ is approaching a continuous
function, $p(x)$ remains discrete, so that the cumulative distribution
of events $P(\geq l)$ is in fact a steeper power law than the density
in this regime, as illustrated numerically in Figure \ref{cuts_figure}.

Next, to compare the continuum model to PLR,  we note that
in PLR, $l(x)$ becomes close to continuous when the
category size $L$ becomes small and the event sizes $l_i$
become very small. Formally, this corresponds to the limit
$L \rightarrow 0$ with $l_i/L \rightarrow 0$ for every
$l_i$ and every $L$.   As for the cuts model, this is a
case where the discrete  PLR model produces a
nearly continuous event size function $l(x)$, although, the
event size probability density $p_i(l_i)$ remains
sufficiently discrete that computing the  cumulative
distribution $P(\geq l_i)$ does not simply correspond to a unit
increase in the exponent, and instead we must be cautious
and  do additional work to compute the cumulative exponent, we did for the PLR
model in Eq.~(\ref{cuts_limit2}) and Appendix A.

\begin{figure}
\resizebox{!}{.25\textwidth}{{\includegraphics{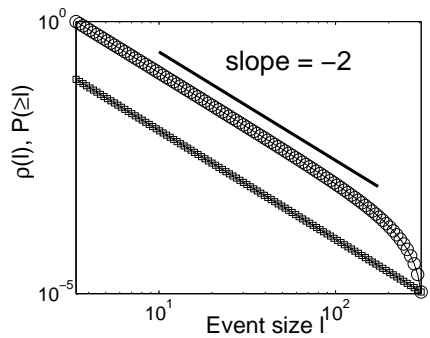} }}
\caption{\label{PLR_small} Plot of probability density $\rho(l_i)$
(small squares), and $P(\geq l_i)$ (large circles) vs. event
size $l_i$ derived from the PLR model. The initial probability density
$p(x)$ is piecewise constant function over intervals of length $L=1$.
$p(x)$ is defined so that the left-hand end-point of each interval has
a value which fits an exponential  density.}
\end{figure}

In $d=1$ PLR predicts that the discrete event size
probability density is $p_i(l_i) \propto l_i^{-2}$, regardless
of the density at which points in that density are sampled.
Furthermore the PLR model begins with the $p_i$ as input
(solving for the $l_i$ by optimizing resource allocations),
so we must work with the density first, then solve for the
cumulative distribution,. Because the $p_i$ and $l_i$ are
discrete, there is no simple relationship between the
density $p_i(l_i)$ and the cumulative distribution $P(\geq l_i)$. 
Naively, one might expect the cumulative probability
to be the integral of the probability density and guess
$P(\geq l_i) \propto l_i^{-1}$. As we have stated
previously, this is emphatically not the case.
Figure~\ref{PLR_small} is a numerical simulation of PLR for
a piecewise constant $p(x)$ (using the  mapping of event
categories into spatial positions,  each of length $L$
given by the width over which $p(x)$ is constant), which is
defined so that the left-hand end-point of each interval
has a value which fits an exponential density function.
This figure shows $\rho(l_i) \propto l_i^{-2}$, as
predicted, yet $P(\geq l_i) \propto l_i^{-2}$ as well. In other
words, the cumulative and noncumulative probabilities on a
log-log plot both have a slope of $-2$.

In fact, it is straightforward to show analytically that
the cumulative slope matches the noncumulative slope in
this case. The PLR probabilities $\{p_i\}$ are given as
exponentially distributed: $p_i \propto e^{-\lambda (i-1)
\; L}$, where $L$ is the category size, which is subdivided
into regions of size $l_i$. The large scale cutoff
$\Lambda$ is $L \times n$, where $n$ is the number of
categories. We can calculate the cumulative probability:
\begin{eqnarray}
P(\geq l_i) &=& \frac{\sum_{j=i}^{n} e^{-\lambda (j-1) \; L} }{\sum_{j=1}^{n} e^{-\lambda (j-1) \; L}} \nonumber 
\\ &=& \sum_{j=i}^{\Lambda/L} e^{-\lambda (j-1) \; L} \Delta(L) \nonumber
\\ &\simeq& \int_{(i-1)\;L}^{\Lambda} e^{-\lambda x} dx  \nonumber
\\ &\simeq& \frac{1}{\lambda} e^{-\lambda (i-1) \; L}\nonumber
\\ &\propto& p_i \propto l_i^{-2} \end{eqnarray} %
where we have used the fact that because $L \rightarrow 0$ the
reciprocal of the norm, $\Delta(L)$, approaches zero and 
we can approximate the sum as an integral. We also drop the
term proportional to $e^{-\lambda \Lambda}$, which is much
smaller than  $e^{-\lambda (i-1) \; L}$. Thus the
cumulative distribution is proportional to the
noncumulative distribution in this limit, and
the continuum, cuts, and PLR models all match in the regime
where resources are dense and event sizes are small.

\subsection{ PLR and cuts for large events}

A final question is whether the PLR model is similar to the
cuts model in the limit of very large event sizes, where
the cuts model predicts $P(\geq l_i)$ has a exponent of
$-1$. As we mentioned earlier, in one dimension the PLR
model predicts $p_i(l_i) \propto l_i^{-2}$ for  every $l_i$
and every $L$. However, cumulative distributions which
result from discrete probability densities can have any one
of a large class of shapes and exponents.  For PLR to
predict a cumulative slope $P(\geq l_i) \propto l_i^{-1}$
(i.e. the same as the cuts model for large events),
the discrete PLR points $\rho(l_i)$ must be sufficiently
dense so that the summation of those points approximates an
integral. This occurs when the $l_i$ increase very slowly,
or equivalently if the spark probability density $p(x)$ is
very heavy tailed. For spark probability densities $p(x)$
(such as the exponential) which drop off quickly, the $l_i$
increase rapidly (see Eq.~(\ref{cuts_recur2})) and PLR will
not predict a slope $\alpha=-1$ for an individual optimized
system.

Interestingly, most data from complex systems like forest
fires~\cite{fire} and web traffic~\cite{PLR} are
sufficiently dense  that an integral approximation is
reasonable. Cumulative slopes of $\alpha=-1/d$ are
consistent with the PLR model when interpreted as in
Section III. As previously mentioned, this might best be
explained by viewing these data sets as {\em mixtures} of
data from systems which are individually optimized.  In
this case a probability density with a sparsely populated
tail (such as Fig.~\ref{cuts_figure}(a)) might be mixed
with similar data so that the tail becomes densely
populated. This is precisely what is done in for the cuts
model in Figure~\ref{cuts_mixture} and here the mixture
power law retains $\alpha=-1/d=-1$. Thus it is possible
that mixtures of PLR models could be made consistent with
the cuts model in the limit of large event sizes. However,
because PLR makes analytic predictions only for
noncumulative probability densities $p_i(l_i)$, in the
absence of a more thorough analysis of mixtures of PLR
solution, we can draw no further general conclusions about
the behavior of cumulative probabilities $P(\geq l_i)$ for
large $l_i$ in this paper. Instead, we reserve this issue
for a more detailed analysis in \cite{mixtures}.

\section{VI. Pathologies of the Lattice Model}

Abstract forest fire models have arisen as paradigms in
complex systems theory, initially  for the SOC mechanism
\cite{SOC_first,FFrefs,FFrefs2} and later also for HOT
\cite{PRE1,PLR, fire}. Inspiration for SOC comes from
statistical physics, where lattice models have played a
central role in theoretical explorations of large scale
consequences of local interactions \cite{Percolation}. HOT
is motivated by biology and engineering, where lattice
models are a less natural starting point. Nonetheless, in
an effort to clarify comparisons between the mechanisms,
and because of their pedagogical explanatory power, study
of HOT also began with lattice models. However, in the
limit of large lattices, the HOT lattice model can  become
somewhat pathological, which led  to the alternative HOT
models analyzed in this paper. In this section we discuss
the nature of this pathology. It arises in what corresponds
to a natural limit for percolation in statistical physics
that goes awry in the analogous HOT model, because the
difference in scaling between the $d-$dimensional
contiguous regions, and the $(d-1)-$dimensional barriers.

SOC builds on the concept of criticality in statistical
physics. The percolation phase transition is associated
with a critical density of occupied sites, at which a
connected cluster of nearest neighbor occupied sites first
spans the lattice (say, from top to bottom) in the limit of
infinite lattice size. Infinitesimally above the critical
density, the infinite cluster exists with probability
converging to unity as the lattice size diverges.
Simultaneously the probability any given site is connected
to the infinite cluster converges to zero. This occurs
because the infinite cluster is a fractal. An immediate
consequence of the fact that the fractal dimension is less
than the lattice dimension is that removal of the infinite
cluster (i.e. in the largest possible fire) does not alter
the lattice density even though the cluster is
system-spanning (i.e. would stretch across the entire
forest). At the critical density, and only at the critical
density, the distribution of cluster sizes in the ensemble
is described by a power law.

In statistical physics power law predictions are typically
sharpened by taking the limit of infinite lattice size.
However, in attempting this for the HOT lattice model a
problem emerges, that makes the large lattice limit
ill-posed.  This also reveals more clearly an intrinsic
flaw in the lattice model when it comes to modeling
mechanisms and costs associated with suppression of fires
and other cascading events in highly designed or evolved
systems \cite{COLD}. Consider the lattice model in $d=2$.
In both the HOT and SOC lattice model a firebreak forms
when any unbroken chain of empty lattice sites isolates a
connected cluster, even if the chain is only one lattice
spacing wide. In SOC (and criticality) the underlying
randomness with which configurations are generated, and the
symmetry between vacant and occupied sites, results in a
critical density of 0.4 (0.59) in $d=2$ which is bounded
away from unity, so that a finite fraction of the lattice
is devoted to both clusters and firebreaks in the limit of
infinite size. In other words, the size of the firebreaks
scales in the same way as the size of the connected
clusters.  However, in the HOT version, simple optimization
of yield (number of trees remaining after a single spark,
averaged over the spark distribution) leads to macroscopic,
compact clusters of trees separated by narrow (one lattice
spacing wide), efficient (linear) firebreaks.  Thus in the
limit of large lattices the cost in density and yield
associated with each firebreak becomes vanishingly small.

To visualize how the cost of firebreaks becomes negligible
for large lattices and why this is a problem, consider
large $N\times N$ lattices as $N\to\infty$. A vertical line
of empty sites extending from top to bottom on the lattice
involves $N$ sites, and so the cost in lattice density
associated with making those sites vacant is $N/N^2=1/N$.
This cut divides an otherwise fully occupied lattice into
two separate regions (left and right of the firebreak). In
the limit $N\to\infty$, the cost in density of the cut is
zero, even though the division of the lattice into two
separate regions is preserved. Similarly, a collection of
equally spaced vertical and horizontal cuts on an otherwise
occupied lattice results in a  gridded configuration
dividing the lattice into square  regions  of equal size,
each outlined by a firebreak one lattice spacing wide on
each of the four sides. For this configuration, all fires
are of equal size (the area of the contiguous square). For
a finite lattice such a solution  could only be optimal for
a spatially uniform distribution of sparks. However, in the
limit $N\to\infty$ an infinite family of such solutions all
achieve the maximum yield of unity. All that is required is
that the cuts be positioned far enough apart that the grid
of firebreaks consume zero density, yet close enough
together that the density cost associated with a fire in
any individual square of contiguous occupied sites  is also
zero. This is achieved whenever  the spacing between grid
lines scales like $N^\gamma$ with $0<\gamma<1$. This
produces a yield of unburnt trees that is asymptotically
perfect (i.e. approaches unity) for the entire forest for
any distribution of sparks, with infinitesimal fire sizes.

It is straightforward to generalize this argument to higher
dimensions, because it relies only on the fact that the
barriers scale differently (like $d-1$) compared to the the
compact regions (like $d$). Unrealistically, a literal
interpretation of the lattice model suggests that with
proper management and minimal cost, essentially all fires
could be eliminated\cite{COLD}. While this form of the HOT
lattice model is useful pedagogically as it exhibits such
striking differences from the SOC version, it has too many
flaws to be taken literally as a model of real forest fires
because the costs of resources for suppression are not
accounted for properly. While there is a natural duality
between vacant and occupied sites in the models of
statistical physics, in HOT models vacancies are resources
which define boundaries that scale differently than the
bulk substrate. For specific applications, resources are
rarely (if ever) simply the absence of substrate.  Even
firebreaks constructed on forest land (e.g. roads) are not
simply the absence of trees, but are cut and maintained at
significant economic expense.

\section{VII. Discussion}

The abstract HOT models studied in this paper correct  the
pathology of the original HOT lattice model by explicitly
accounting for resource use. The PLR and cuts models do
this through an explicit cap on the total resources
available. The continuum model does this through inclusion
of an explicit resource cost term in the yield function.
Several preliminary calculations suggest that at least
within a range of functional representations, the specific
manner in which resources are accounted for  is not a
crucial  factor in determining the exponent in the power
law for these models. For example, a more general
cost-benefit term describing resource use can replace the
explicit cap on resources in the PLR model,  at the expense
of analytical tractability of the model, but with no
significant change in the  exponent. Analogously, the cuts
model (in the limit of small event sizes) and the continuum
model can  lead to the same power law exponent, in spite of the
fact that they account for the cost of resources in
different ways.

The key feature in determining the size distribution for a
given model is that we optimize, while measuring the cost (or
loss) in terms of the average event size. Alternative
formulations of the continuum model \cite{PRE1,COLD} have
considered alternative cost or utility functions, which
clearly can lead to modifications in the event size
distribution. For example, if the cost function puts  a
large penalty for events greater than a given size, then
more resources will be devoted to large events, at the
expense of more smaller events, and a great average size.
Such considerations are clearly relevant in cases such as
finance and economics, where risk-seeking and risk-averse
strategies come into play.

Compared with models based on criticality, the power laws
predicted by all of the HOT models are much steeper, and
have the opposite trends with dimensionality. In
criticality the exponents become  smaller for lower
dimensional problems. This is the opposite of the trends
observed in data \cite{PLR}, which typically  exhibit
steeper power laws for lower dimensional problems, as in
HOT. It is worth noting (especially given our focus on
$d=1$) that while percolation in $d=1$ has the (trivial)
critical density of unity-- the only way connectivity can
arise across a one dimensional lattice is for every site to
be occupied-- the configurations and size distribution
(not a power law in $d=1$) which
arise in random percolation in the neighborhood of the
critical density even in that case are completely unlike
those that arise in the corresponding one-dimensional HOT
lattices. In criticality, the  placement of vacancies is
random, whereas in HOT the specific placement of vacancies
is dictated by  optimization.

In models based on criticality, the self-similar, fractal
event shapes, reflect a mechanism which is intrinsically
scale-free, producing a single exponent, spanning all
scales. In contrast, in HOT models  heavy-tailed
distributions  arise from optimization on a macroscopic scale.
Compact regions predicted by HOT are not fractal or
self-similar and there is no reason to expect that small
scale events will {\it a priori} be described by the same
power law as large scale events.

The cuts model is a clear example in which we do observe a
heavy-tailed event size distribution, with asymptotically
different power law exponents as we vary the scale. This
model highlights the essential difference between the dense
and sparse resource regimes, which in the original
formulations of the continuum and  PLR  models emerge from
the distinction between inherently continuum and discrete
fields describing probabilities, resources, and losses. In
the continuum case, it is simply not possible to capture
features which could arise as a consequence of discrete,
sharp, well-separated boundaries-- the sparse resource
regime. Thus  the continuum model agrees with the cuts
model only in the limit that the cuts (which are sharp and
discrete) are placed asymptotically close together, i.e.
the dense resource limit. On the other hand, the PLR model,
which assumes discrete event categories, can in principle
capture both the dense resource limit and the sparse
resource regime, though the latter will need additional
treatment because of the intrinsic role that mixtures play
in real data. In this paper we explored the PLR model in
the limit of dense resources, by taking the length scales
of the system $L$ and the event sizes $l_i/L$
simultaneously to zero. In this limit, the PLR model can
capture the the continuous, spatial spark distribution
$p(x)$, though PLR (and cuts) remain intrinsically
discrete.

Based on this analysis, it may appear  that the cuts model
is the clear winner, simultaneously capturing the full
range of behaviors seen in the other two, and this would be
true if we only considered $d=1$. However, in order to
generalize the cuts model beyond $d=1$ it is necessary to
constrain the optimization procedure. For example, in
\cite{PRE1} this was done by specifying a grid design. In
many cases, such a constrained design may not be desirable,
and the abstractions of the other models may be preferred.
The continuum and  PLR models are both easily formulated in
arbitrary dimension $d$, but with different predictions for
the exponents. As we've shown here, the PLR model can be
extended to the dense resource regime, where it agrees with
the predictions of the continuum model. The reverse is not
the case. In that sense, the continuum model is less
flexible. Furthermore, the PLR model has been far more
successful in capturing statistics of event size
distributions, assuming data sets are dense enough to be
described as continuous distributions (e.g. assuming they
are mixtures \cite{mixtures}).  Examples which have been
studied include world wide web traffic, forest fires, and
power outages \cite{PLR,CORNELL,PNAS}.

In comparison, we do not yet have any clear examples where
the predictions of the continuum model have been shown to
apply. Perhaps the reason behind this lies in the fact that
data is almost exclusively collected for large events in
the sparse resource regime. In regimes where resources are
abundant, one may simply choose not to optimize. Small file
downloads, fires, and outages are rarely monitored, and
small scale cutoffs, whether deliberately imposed for
convenience or arising from  an inherent physical
mechanism, tend to prevent detailed statistical analysis of
this regime. In any case, statistical distributions remain
only a starting point for understanding mechanisms for
complexity and modeling system failure. Success arises from
the study of simple models when their predictions capture
aspects of the system which can be described and quantified
at a relatively low resolution. From this initial success,
they can inspire a sequence of higher resolution models and
observations to understand and anticipate detailed
mechanisms for cascading failure in natural and
technological systems.

\appendix
\section{Appendix A: Asymptotic limits of the cuts model for a
power law initial probability density}
In this appendix we derive the slope of $P( \geq l_i)$ on a
log-log plot for a cuts model where the initial probability
density is described by a power law, $ p(x) = a x^{-(a+1)}$. First
we use the cuts model to find an analytic description for the set
of discrete probabilities $p_i$ and event sizes $l_i$:
\begin{eqnarray}
p_i &=& (c_{i-1})^{-a} - (c_{i})^{-a} \nonumber \\
l_i &=& \frac{p_i}{p_(c_i)} \nonumber \\
l_i &=& \frac{(c_{i-1})^{-a} - (c_{i})^{-a}}{a(c_i)^{-(a+1)}}
\label{cuts_recur_power_2}
\end{eqnarray}

We also recall the definitions for the cuts positions $c_i$ and
the cumulative probabilities $p^{t}_{i} =P( \geq l_i)$:

\begin{eqnarray}
c_i &=& c_{i-1} + l_i \nonumber \\
p^{t}_{i} &=& P( \geq l_i) =  (c_{i-1})^{-a}
\end{eqnarray}

The slope of $P( \geq l_i)$ on a log-log plot can be
calculated in limiting cases by  dividing $\Delta \log
p^{t}_{i}$ by $\Delta \log l_{i}$, Taylor expanding and dropping
higher order terms.

\begin{eqnarray}
\label{power_step2}
\frac{\Delta \log
p^{t}_{i}}{\Delta \log l_{i}} = \frac{\log p_{i+1}^{t}-\log
p_{i}^{t}}{\log l_{i+1}- \log l_{i} } \nonumber \\
= \frac{-a \log(c_{i-1} + l_{i}) + a \log(c_{i-1})}
{\log \left[\frac{ c_{i-1}^{-a} - (\cut + l_i)^{-a}}
{a( \cut + l_i)^{-(a+1)}}\right] - \log l_i}
\end{eqnarray}

Now we will assume that $l_i$ is small compared to $\cut$ and we
will  derive terms which can be Taylor expanded to first
order in $\frac{l_i}{\cut}$. We first evaluate numerator of
Eq.~(\ref{power_step2}):

\begin{eqnarray}
\lefteqn{-a \log(c_{i-1} + l_{i}) + a \log(c_{i-1})} \nonumber \\
&=& a \log \cut - a \log(1 + \frac{l_i}{\cut}) + a \log(\cut)
\nonumber \\
&=& -a \log(1 + \frac{l_i}{\cut})
\label{numerator}
\end{eqnarray}

Now we evaluate the denominator:

\begin{eqnarray}
\lefteqn{\log \left[\frac{ c_{i-1}^{-a} - (\cut + l_i)^{-a}}{a(
\cut + l_i)^{-(a+1)}}\right] - \log l_i} \nonumber \\
&=&  -\log a - \log l_i + (a+1) \log( \cut + l_i) - a \log \cut
\nonumber \\
& &   + \log\left(1-(1+ \frac{l_i}{\cut})^{-a}\right) \nonumber
\\
&=& -\log(a l_i) + (a+1)\log \cut + (a+1) \log\left(1 +
\frac{l_i}{\cut}\right) \nonumber \\
& &    -  a \log \cut + \log\left(1-\left(1+
\frac{l_i}{\cut}\right)^{-a}\right)
\label{denominator}
\end{eqnarray}
We assume $l_i/ \cut <\!< 1$ and use the binomial expansion on
the last term.  Then Eq.~(\ref{denominator}) becomes:
\begin{eqnarray}
\lefteqn{\log \left[\frac{ c_{i-1}^{-a} - (\cut + l_i)^{-a}}{a(
\cut + l_i)^{-(a+1)}}\right] - \log l_i} \nonumber \\
&=& \log\left(\frac{\cut}{a l_i}\right) + (a +1)\log \left(1 +
\frac{l_i}{\cut}\right) \nonumber \\
& &  + \log\left(1 - \left(1- \frac{a l_i}{\cut} +
\frac{a(a+1)}{2}(\frac{l_i}{\cut})^{2}\right)\right) \nonumber \\
&=& \log\left(\frac{\cut}{a l_i}\right) + (a +1)\log \left(1 +
\frac{l_i}{\cut}\right) + \nonumber \\
& & \log\left(\frac{a l_i}{\cut}\left(1 -
\frac{(a+1)}{2}(\frac{l_i}{\cut}\right)\right) \nonumber \\
&=& (a+1)\log\left(1 + \frac{l_i}{\cut}\right) + \log\left(1 -
\frac{(a+1)l_i}{2\cut}\right)
\label{denominator2}
\end{eqnarray}
Inserting the numerator and denominator back into
Eq.~(\ref{power_step2}) we have:
\begin{eqnarray}
\lefteqn{\frac{\Delta \log p^{t}_{i}}{\Delta \log l_{i}}} \nonumber \\
&=&\frac{-a \log\left(1 + \frac{l_i}{\cut}\right)}{(a+1)\log\left(1
+ \frac{l_i}{\cut}\right) + \log\left(1 -
\frac{(a+1)}{2}(\frac{l_i}{\cut})\right)}
\label{power_step8}
\end{eqnarray}
Now we use the Taylor expansion $\log(1+ \epsilon)= \epsilon +
{\cal O} (\epsilon^{2})$:
\begin{equation}
\frac{\Delta \log p^{t}_{i}}{\Delta \log l_{i}}
\simeq \frac{-a \frac{l_i}{\cut}}{(a+1)(\frac{l_i}{\cut}) -
\frac{(a+1)}{2}(\frac{l_i}{\cut})}
\end{equation}
\begin{equation}
\frac{\Delta \log p^{t}_{i}}{\Delta \log l_{i}} = \frac{-2
a}{a+1}
\end{equation}
This is the slope of $P( \geq l_i)$ on a log-log plot in the
limit where $l$ becomes small.

Now we will look in the opposite limit, where $l$ becomes large.
We first show that
$l \rightarrow \infty$ implies $\frac{\cut}{l_i}
\rightarrow 0$ if $a>1$. Using the definition for $l_{i+1}$
in Eq.~(\ref{cuts_recur_power_2}) we derive a recursion  relation
for $c_i / l_{i+1} \equiv g_i$.

By definition:
\begin{eqnarray}
\lefteqn{g_i \equiv \frac{c_i}{l_{i+1}}} \nonumber \\
 &=& \frac{ a\; c_i\; (c_i)^{-a}(c_i)^{-1}}{(\cut)^{-a} - (c_i)^{-a}} \nonumber \\
 &=& \frac{a \left( \cut + l_i \right)^{-a}}{(\cut)^{-a} - \left(\cut + l_i\right)^{-a}} \nonumber \\
 &=& \frac{a \: l_{i}^{-a}\left(\frac{\cut}{l_i}+1\right)^{-a}}{l_{i}^{-a}\left(\left(\frac{\cut}{l_i}\right)^{-a} - \left(\frac{\cut}{l_i}+1\right)^{-a}\right)} \nonumber \\
 &=& \frac{a \left(g_{i-1}+1\right)^{-a}}{\left((g_{i-1})^{-a} - \left(g_{i-1}+1\right)^{-a}\right)}
\end{eqnarray}
We note that $g_i$ will always be less than 1. Therefore we can use the binomial
expansion and write out the terms to lowest order in $g_{i-1}$:
\begin{equation}
g_i= \frac{a \left(1 - a g_{i-1}+ {\cal O}(g_{i-1}^{2})\right)}{(g_{i-1})^{-a} - \left(1- a g_{i-1}+ {\cal O}(g_{i-1}^{2})\right)}
\label{g_middle}
\end{equation}
Now we note that if we assume $g_{i-1} << 1$ for large $i$ we can drop all terms
of order $g_{i-1}^{2}$. Also, for $a>1$ the first term in the denominator will be
much larger than the other terms, and we drop all the other terms in the
denominator. Then we have:
\begin{equation}
g_i \simeq \frac{a \; (1 - a g_{i-1})}{(g_{i-1})^{-a}} \simeq a(g_{i-1})^{a} + a^{2}(g_{i-1})^{a+1}
\end{equation}
We can then find the ratio of consecutive terms:
\begin{equation}
g_i/g_{i-1}  \simeq a\;(g_{i-1})^{a-1} + a^{2}(g_{i-1})^{a}
\end{equation}
Because $g_i<1$ for all $g_i$, we see that our assumption that
$g_i << 1$ was indeed valid, and that the sequence goes to zero
as $i$ approaches infinity.  We also note that if $a<1$, we can
no longer assume that the first term in the denominator in
Eq.~(\ref{g_middle}) is much larger than 1.  In fact, as
$a \rightarrow 0$ the first term in the denominator approaches 1,
and it is not true that $\frac{\cut}{l_i}$ approaches zero for
large $l_i$.

We now solve  Eq.~(\ref{power_step2}) for terms
which we can Taylor expand to first order in $\frac{\cut}{l_i}$.
First we simplify the numerator:
\begin{eqnarray}
\lefteqn{-a \log(c_{i-1} + l_{i}) + a \log (c_{i-1}) } \nonumber
\\
&=& -a \log l_i - a \log \left(\frac{\cut}{l_i} +1 \right) + a
\log \cut \nonumber \\
&=& a \log \left( \frac{\cut}{l_i} \right) - a \log
\left(\frac{\cut}{l_i} +1 \right) \nonumber \\
&\simeq& a \log \left( \frac{\cut}{l_i} \right) + \left\{ -a
\left( \frac{\cut}{l_i} \right) + {\cal O} \left(
\frac{\cut}{l_i} \right)^{2} \right\}
\label{numerator2}
\end{eqnarray}
where in the last line we have used the Taylor expansion $\log(1+
\epsilon)= \epsilon + {\cal O} (\epsilon^{2})$.
As  $\frac{\cut}{l_i}$ approaches $0$, $\log
\frac{\cut}{l_i}$ becomes large and negative, and the terms
inside the braces in Eq.~(\ref{numerator2}) become negligible.
Therefore the numerator in this limit is:
\begin{equation}
\Delta \log
p^{t}_{i} \simeq a\log \left( \frac{\cut}{l_i} \right)
\label{numerator_last}
\end{equation}
We simplify the denominator of  Eq.~(\ref{power_step2}).
\begin{eqnarray}
\lefteqn{\log \left[\frac{ c_{i-1}^{-a} - (\cut + l_i)^{-a}}{a(
\cut + l_i)^{-(a+1)}}\right] - \log l_i} \nonumber \\
&=& -\log(a l_i) - (a+1) \log \left(\frac{\cut}{l_i} +1 \right)
\nonumber \\
& & + (a+1) \log l_i - a \log l_i \nonumber \\
& & + \log \left( \left( \frac{\cut}{l_i} \right)^{-a} -
\left(\frac{\cut}{l_i} +1 \right)^{-a}\right)
\label{denominator3}
\end{eqnarray}
Again we use the binomial expansion to approximate the last term
in the denominator:
\begin{eqnarray}
\lefteqn{ \log \left( \left( \frac{\cut}{l_i} \right)^{-a} -
\left(\frac{\cut}{l_i} +1 \right)^{-a}\right) } \nonumber \\
&=& \log \left( \left( \frac{\cut}{l_i} \right)^{-a} - \left(1 -
\frac{a \cut}{l_i} + a(a+1)  \left( \frac{\cut}{l_i} \right)^{2}
\right) \right) \nonumber \\
&\simeq& \log \left( \left( \frac{\cut}{l_i} \right)^{-a} \right)
\end{eqnarray}
where in the last line we have used that $(\frac{\cut}{l_i})^{-a}
>\!> 1 > (\frac{\cut}{l_i})$. Then the denominator
(Eq.~\ref{denominator3}) becomes
\begin{eqnarray}
\lefteqn{\log \left[\frac{ c_{i-1}^{-a} - (\cut + l_i)^{-a}}{a(
\cut + l_i)^{-(a+1)}}\right] - \log l_i} \nonumber \\
&=& - \log a + (a+1)\log \left(\frac{\cut}{l_i} +1 \right) + \log
\left( \left( \frac{\cut}{l_i} \right)^{-a} \right) \nonumber \\
&=& - a \log \left( \frac{\cut}{l_i} \right) + (a+1)\log
\left(\frac{\cut}{l_i} +1 \right) - \log a \nonumber \\
&=& - a \log \left( \frac{\cut}{l_i} \right) + \nonumber \\
& &\left\{ -(a+1)\left[ \frac{\cut}{l_i} + {\cal O} \left(
\frac{\cut}{l_i} \right)^{2} \right] - \log a \right\}
\label{denominator4}
\end{eqnarray}
where in the last line we used the Taylor expansion for $\log(1 +
\epsilon)$. Again we see that for any finite $a$ the terms inside
the braces in the last line of Eq.~(\ref{denominator4}) become
negligible compared to $\log \left( \frac{\cut}{l_i} \right)$ in
the limit that $\frac{\cut}{l_i}$ becomes small. In this limit
the denominator can be approximated as:
\begin{equation}
\Delta \log l_{i} \simeq - a\log \left( \frac{\cut}{l_i} \right)
\label{denominator5}
\end{equation}
Dividing Eq.~(\ref{numerator_last}) by Eq.~(\ref{denominator5}),
we see that for small $\frac{\cut}{l_i}$, the slope of $P(\geq l)$ on
a log-log plot is:
\begin{equation}
\frac{\Delta \log p^{t}_{i}}{\Delta \log l_{i}} = -1
\end{equation}
We have shown that if $a$, the exponent for the power law spark 
distribution, is greater than 1, then $\frac{\cut}{l_i}$ becomes
small as $l_i$ becomes large. In this case we have shown $-1$ to be the
asymptote of the exponent of $P(\geq l)$  for large $l$.

\section{Appendix B: Optimal solutions for two different cost
functions}
We are interested in comparing the optimal solutions
for two different cost functions (equivalently, for two different
yield functions
$Y$ in Eq.~(\ref{cuts_yield}) and $Y^t$ in
Eq.~(\ref{cuts_yield2})). The first cost
function $J = \sum p_i l_i$ equates cost with expected
event size, and is used in PLR and continuum models.
The second cost function $J^t = \sum p_{i}^{t} l_i$
equates cost with expected {\em transferred} event size
and is used in the cuts model. This situation arises
when the frequency with which an event is
``transferred'' is equal to the cumulative probability
of all larger events. One example is sequentially
linked web files. Though $J^{t}$ is less intuitive than
$J$, it has the very nice property that one can
analytically solve for the optimal event sizes $l_i$
given cost function $J^{t}$. In most situations,
however, we are really interested in optimizing the
original cost function $J$.

 In this section we will show that the optimal
solutions $\{l_i\}$ are the {\em same} for either
definition of cost ($J^{t}$, $J$) in the limits $l_i
\rightarrow \infty$ and $l_i \rightarrow 0$. This
allows us to directly compare analytic results from the
cuts model with results from continuum and PLR models
in limiting cases.

First, we recall that optimizing $J$ leads to a
recursion relation for optimal event sizes $l_i$,

\begin{equation}
\label{cuts_recur1_app}
  p_i + \left[l_i p(c_i) - p_{i+1}\right] = l_{i+1} \: p(c_i)
\end{equation}

while optimizing $J^{t}$ leads to a different recursion
relation.

\begin{equation}
\label{cuts_recur2_app}
  p_i = l_{i+1}\: p(c_i)
\end{equation}

Comparing Eq.~(\ref{cuts_recur2_app})
and~(\ref{cuts_recur1_app}), we see they give the same
result if the bracketed term in Eq.~(\ref{cuts_recur1_app}),
$\epsilon \equiv l_i p(c_i) - p_{i+1}$, is much
smaller than $p_i$. First we will show this is the the
case in the limit $l_i \rightarrow 0$.

We want to show:
\begin{eqnarray}
\label{cuts_start_app}
\epsilon = l_{i} p(c_i) - p_{i+1} & <\!< & p_i
\nonumber \\
l_{i} p(c_i)  & <\!< & p_i + p_{i+1}
\end{eqnarray}

We can rewrite the right-hand side as

\begin{equation}
\label{cuts_pavg_app}
p_i + p_{i+1} = \int_{c_{i-1}}^{c_{i+1}} p(x) dx =
p_{avg}(l_{i+1} + l_i)
\end{equation}

where $p_{avg}$ is the average value of $p(x)$ on the
interval [$c_{i-1}$,$c_{i+1}$].

Then Eq.~(\ref{cuts_start_app}) can be rewritten as

\begin{equation}
\label{cuts_zero_app}
(p(c_i) - p_{avg}) l_i <\!< p_{avg} \; l_{i+1}
\end{equation}

We use the recursion relation for the event sizes given
by Eq.~(\ref{l_iter}):

\begin{equation}
\label{cuts_recurl_app}
l_{i+1} = \frac{e^{\lambda l_i} -1}{\lambda}
\end{equation}

In the limit $l_i \rightarrow 0$ this implies $l_{i+1}
= l_i + {\cal O}(l_i^2)$. Neglecting terms of order
$l_i^2$ we have

\begin{equation}
\label{cuts_last_app}
\left(p(c_i) - p_{avg}\right) <\!< p_{avg}
\end{equation}

The position $c_i$ approaches the midpoint of the
interval [$c_{i-1}$,$c_{i+1}$] because $l_{i+1}
\rightarrow l_{i}$. As the length of the interval goes
to 0, the value of $p(x)$ at the midpoint, $p(c_i)$,
approaches the average value of $p(x)$ over the
interval. Therefore the left hand side of
Eq.~(\ref{cuts_last_app}) {\em is} negligible compared
to the right-hand side. In the limit $l_i \rightarrow
0$,  $\epsilon$ is much smaller than $p_i$.

Now we will show that this is the case for the limit as
$l_i \rightarrow \infty$. Starting from
Eq.~(\ref{cuts_start_app}) we again want to show:

\begin{equation}
\label{cuts_start2_app}
l_i \;p(c_i) <\!< p_{i+1} + p_{i} =
\int_{c_{i-1}}^{c_{i+1}} p(x) dx
\end{equation}

Substituting $\lambda e^{- \lambda x}$ for $p(x)$ we
have

\begin{eqnarray}
\label{cuts_sub_app}
l_i \lambda e^{ -\lambda c_i} &<\!<& - e^{- \lambda
c_{i+1}}+ e^{- \lambda c_{i-1}} \nonumber \\
 &<\!<& e^{- \lambda c_{i-1}}(e^{- \lambda (l_i +
l_{i+1})} + 1) \nonumber \\
 &<\!<& e^{- \lambda c_{i-1}}
\end{eqnarray}

where we have used $e^{- \lambda l_i} <\!< 1$. Then we
have

\begin{eqnarray}
\label{cuts_last2_app}
\log( \lambda l_i) - \lambda c_i &<\!<& -\lambda
c_{i-1} \nonumber \\
\log( \lambda l_i) & <\!< & \lambda l_i \nonumber \\
\end{eqnarray}

For $l_i \rightarrow \infty$ and $\lambda$ fixed,
$\log( \lambda l_i)$ {\em is} negligible compared to
$\lambda l_i$. Therefore in the limit $l_i \rightarrow
\infty$,  $\epsilon $ is much smaller than $p_i$.

Therefore in these two limits, optimizing $J^{t}$
results in the same optimal event sizes as optimizing
$J$.

\vskip .1truein

{\it Acknowledgements:} This work was supported by the
David and Lucile Packard Foundation, NSF Grant No.
DMR-9813752, the James S. McDonnell Foundation, and
the Institute for Collaborative Biotechnologies through
grant DAAD19-03-D-0004 from the U.S. Army Research Office.
M.M. was supported by a National Science Foundation
Graduate Research Fellowship.


\begin{thebibliography}{999}
\bibitem{a_2} For $a>2$,
$p(x)$ has finite variance, i.e. a \lq\lq light tail," and
is no longer a stable law.
\end{thebibliography}
\end{document}